\input ppltex.sty
\input abbrev.sty

\rjustline{PPPL--1528 (Apr.~1979)}
\rjustline{Phys.\ Fluids \vol{22}(11), 2188--2209 (Nov.~1979)}
\vskip 0.5in
\ctrline{\titl Stochastic Ion Heating by a Lower Hybrid Wave: II}
\vskip 20 pt
\ctrline{Charles F. F. Karney}
\vskip 5 pt
\ctrline{Plasma Physics Laboratory, Princeton University,}
\ctrline{Princeton, New Jersey 08544}

\abstract

The motion of an ion in a coherent lower hybrid wave (characterized
by $\abs{k_\para} \lsls \abs{k_\perp}$ and $\omega \grgr \Omega_i$)
in a tokamak plasma is studied.  For ions satisfying $v_\perp
> \omega/k_\perp$, the Lorentz force law for the ions is
reduced to a set of difference equations which give
the Larmor radius and phase of an ion on one cyclotron
orbit in terms of these quantities a cyclotron period
earlier.  From these difference equations an earlier result
[Phys.\ Fluids \vol{21}, 1584 (1978)] that above a certain
wave amplitude the ion motion is stochastic, is readily
obtained.  The stochasticity threshold is given a simple physical
interpretation.  In addition, the difference equations are used to derive
a diffusion equation governing the heating of the ions above
the stochasticity threshold.  By including the effects of
collisions, the heating rate for the bulk ions is obtained.

\section{I}{Introduction}

The motion of an ion in a plasma in the presence of a lower
hybrid wave becomes stochastic if the amplitude of the lower
hybrid wave exceeds a threshold.  This provides a mechanism
by which ions may be directly heated by the wave when using lower
hybrid waves to heat a tokamak plasma.  In this paper we extend
the work of an earlier paper\ref{1} (henceforth referred to as
I), in which the stochasticity threshold was derived.  The main
objects are to give a simple physical explanation of the stochasticity
threshold, to derive the velocity space diffusion coefficient,
and hence to determine the heating rate.

In I it was shown that for the purposes of computing the ion
motion it may be assumed that the magnetic field is uniform
if the fractional change in the magnetic field over a Larmor radius
is small.  Similarly, the lower hybrid wave may be treated as
perpendicularly propagating if the wavevector, $\vec{k}$,
satisfies $\abs{k_\para/k_\perp} < \half (\omega/\Omega_i)
^{-1/6}$.  Both these conditions are well satisfied in normal
circumstances.  Shear in the magnetic field may also affect
the motion;\ref{2}  however, if the scale length of the shear
greatly exceeds the Larmor radius, we expect that this effect may also
be ignored.  Thus, in this paper, we consider only
the case of an electrostatic wave propagating perpendicularly to
a uniform magnetic field;  i.e.,
$$\vec{B} = B_0\^{\vec{z}},\qquad
\vec{E} = E_0\^{\vec{y}} \cos(k_\perp y - \omega t). \eqn{\en{fields}{1}}$$
\xdef\fields{\eqprefix {1}}%
As in I these fields are taken to be imposed, since only
the motion of a small fraction of the ions on the tail of the
ion distribution function becomes stochastic, while the bulk ions and
electrons support the wave.  Having determined the heating rate it will
be possible to allow for the damping of the wave.

Following I the Lorentz force law for an ion with mass,
$m_i$, and charge, $q_i = Z_i e$, may be written as
$$\" y + y = \a \cos(y - \nu t),\qquad
\.x = y, \eqn{\en{lfl}{2}}$$
\xdef\lfl{\eqprefix {2}}%
where length is normalized to $k_\perp^{-1}$ and time to
$\Omega_i^{-1}$ ($\Omega_i = q_i B_0 / m_i$) and where
$$\eqalignno{\nu &\eqv \omega/\Omega_i, &(\en{pdef}{3}a)\cr\eqskp
	\a &\eqv {E_0/B_0\over \Omega_i/k_\perp}. &(\+b)\cr}$$
\xdef\pdef{\eqprefix {3}}%
For later use we shall also define a normalized Larmor radius,
$r$, by
$$r \eqv k_\perp v_\perp / \Omega_i, \eqn{\+c}$$
where $v_\perp^2 = \.x^2 + \.y^2$.

The plan of this paper is as follows:  We first reduce the Lorentz
force law for the ion, (\lfl), to a set of difference
equations (Sec.\ II) which determine the velocity and phase of the
ion in terms of its velocity and phase a cyclotron period
earlier.  These difference equations have several symmetries which are
presented in Sec.\ III and which simplify the study of the ion
motion.  This is followed by a study of the ion motion for
infinitesimal $E_0$ (Sec.\ IV) where we show that the motion when
$\omega$ is at a harmonic or half harmonic of the cyclotron frequency
has a very different character from the motion at other frequencies.  In
Sec.\ V we describe qualitatively the motion when $E_0$ is finite.  Here
we define the stochasticity threshold and
explain it physically.  The statisical properties of the ion
motion above the stochasticity threshold are described by
the Krylov--Kolmogorov--Sinai entropy and the correlation function
which are examined in Secs.\ VI and VII\null.  From the correlation
function we find the diffusion coefficient (Sec.\ VIII).  For
times much longer than the correlation time the ion motion in
perpendicular velocity space is governed by a diffusion equation
which is derived in Sec.\ IX\null.  This equation is checked
against the exact equations of motion by comparing
a Monte Carlo solution of the diffusion equation with a simulation
described in I\null.  Since the stochastic heating only affects the
tail ions we include a collision term in the diffusion equation as
a means of heating the bulk ions and electrons.  The
resulting {\FP} equation which describes the ion motion in
{\Twod} velocity space ($v_\perp$ and $v_\para$) is reduced
to a {\Oned} equation in $v_\perp$ only (Sec.\ X).  The {\Oned}
{\FP} equation is checked against the {\Twod}
equation by numerically solving each of them
(Sec.\ XI).  Analytic steady-state solutions to the {\Oned}
equation
are derived in Sec.\ XII which give results for the heating
rates of the ions and electrons.

\section{II}{Derivation of the Difference Equations}

In this section we derive difference equations which
approximately describe the motion of the ion.  The procedure involves
integrating the equations of motion along unperturbed orbits.  Nonlinearity
is introduced because we repeatedly correct the unperturbed
trajectories.  We assume
that the wave frequency is much larger than the cyclotron
frequency;  i.e., $\omega \grgr \Omega_i$ or $\nu \grgr 1$.  (This
is the case for lower hybrid waves.)  On a time scale between the
wave period and the cyclotron period the ion does not experience
the effects of the magnetic field and behaves, in some respects,
as though in zero magnetic field.  Thus, when it passes through the
wave-particle resonance points, $\omega = \vec{k}\dot\vec{v}$
or $\.y = \nu$ (Fig.\ 1), it will experience kicks, which
we can approximate by impulses.  This is the physical
picture that will guide our derivation.  We shall not be too concerned
yet about the limits in which our approximations are valid,
prefering to rely on physical intuition.  Appendix A gives an
alternate derivation starting from the Hamiltonian.  While this
derivation is less transparent, it does enable us to state the limits
of validity of the difference equations.

From Fig.\ 1 we see that for $r > \nu$ the particle passes
through wave-particle resonance twice per cyclotron orbit.  We will
approximate the force on the particle due to the
electric field by impulses or delta functions at these two points
and zero elsewhere.

In order to evaluate the area of the delta function forces
we Taylor-expand the orbit about the wave-particle resonance point
using the unperturbed equation of motion, $\" y + y = 0$.  Thus,
we have
$$\eqalignno{y - \nu t &= y_c + \.y(t_c)t^\prime + 
\half \" y(t_c)t^{\prime 2} - \nu t\cr\eqskp
&= \phi_c - \half y_c t^{\prime 2}, &(\e)\cr}$$
where $t^\prime = t - t_c$, $\phi_c = y_c - \nu t_c$,
$t_c$ and $y_c$ are the time and position of the wave-particle
``collision,'' and, by assumption, $\.y(t_c) = \nu$.  We
substitute (\+) into the
electrostatic force term in (\lfl), $\a \cos(y - \nu t)$,
and approximate the resulting expression by $B\,\delta(t-t_c)$,
where $\delta$ is the Dirac delta function and
$$\eqalignno{B &= \int_{-\inf}^\inf \a \cos(\phi_c - \half y_c t^{\prime 2})
dt^\prime \cr\eqskp
&= \a (2\pi/\abs{y_c})^{1/2} \cos[\phi_c - \sign(y_c)\pi/4]
&(\en{Beq}{5})\cr}$$
\xdef\Beq{\eqprefix {5}}%
[see Ref.\ 3, Eq.\ (4.3.144)].  We
note that $\phi_c$ and hence $B$ are the same (to
first order in $\a$) whether we compute $t_c$ and $y_c$ from the
intersection of the orbit for $t < t_c$ with $\.y = \nu$
or from the intersection of the orbit for $t > t_c$ with $\.y =
\nu$.  This is so because $y_c$ and $t_c$ are
unchanged by the kick the ion receives, and because
$d(y - \nu t)/dt = 0$ near the collision (since $\.y = \nu$).  As
a consequence the mapping we define with the difference equations
will be invariant to time reversal, which in turn means that the
mapping is measure-preserving, a necessary condition for a
non-dissipitive (i.e., Hamiltonian) system.

We define the cyclotron orbit as beginning and ending at
$y = 0$, $\.y < 0$ (Fig.\ 1).  Subscript, $j$, will refer to the
orbit at the beginning of the $j$th or ending of the $(j - 1)$th
orbit.  A subscript, $j + \half$, will be used to refer to the middle
of the $j$th orbit, where $y = 0$, $\.y > 0$.  Finally,
subscripts, $+$ and $-$, will refer to the collisions with
$y_c > 0$ and $y_c < 0$.

We describe the orbit by two parameters, a reduced Larmor radius,
$$\r \eqv (r^2 - \nu^2)^{1/2} - \nu \cos^{-1}(\nu/r) + \nu\pi -
\pi/4, \eqn{\en{vdef}{6}a}$$
\xdef\vdef{\eqprefix {6}}%
and the phase of the wave,
$$\t \eqv \nu t \mod{2\pi}.\eqn{\+b}$$
(Note that for $r \grgr \nu$, $\r \sim r$.)  It is convenient to
introduce the sum and difference of these quantities,
$v = \t + \r$, $u = \t - \r$.

Evaluating $B_-$ we note that $\phi_c = - (r_j^2 - \nu^2)^{1/2}
- \nu [t_j + \pi - \cos^{-1}(\nu/r_j)] = - v_j - \pi/4$, so
that
$$B_- = \a (2\pi)^{1/2} (r^2 - \nu^2)^{-1/4} \cos v_j . \eqn{\e}$$
In (\+) we have intentionally dropped the subscripts on $r$
in the amplitude term.  The justification for this is that
$r$ is large, so a change of order $\a$ in $r$ has negligible
effect on the amplitude factor compared with the same
change in the argument to the cosine.

Working to order $\a$ we then find
$$\eqalignno{r_{j + 1/2} &= r_j + (\nu/r)B_-,& (\e a)\cr\eqskp
t_{j + 1/2} &= t_j + \pi - {(r^2 - \nu^2)^{1/2}\over r^2} B_- ,&
(\+b)\cr}$$
whence from (\vdef)
$$\eqalignno{\r_{j + 1/2} &= \r_j + \nu(r^2 - \nu^2)^{1/2} r^{-2} B_- ,&
(\e a)\cr\eqskp
\t_{j + 1/2} &= \t_j + \nu\pi
- \nu (r^2 - \nu^2)^{1/2} r^{-2} B_- & (\+b)\cr}$$
[note that $\pd\r/\pd r = (r^2 - \nu^2)^{1/2}/r$], and
from (\Beq)
$$\eqalignno{u_{j + 1/2} &= u_j + \nu \pi - 2\pi A \cos v_j ,&
(\en{foo}{10}a)\cr\eqskp
v_{j + 1/2} &= v_j + \nu\pi ,& (\+b)\cr}$$
\xdef\foo{\eqprefix {10}}%
where
$$A = \left( {2\over \pi} \right) ^{1/2}
{\a\nu(r^2 - \nu^2)^{1/4}\over r^2} . \eqn{\en{Adefp}{11}}$$
\xdef\Adefp{\eqprefix {11}}%
Similarly, we may show that
$$\eqalignno{u_{j + 1} &= u_{j + 1/2} + \nu\pi ,& (\e a)\cr\eqskp
v_{j + 1} &= v_{j + 1/2} + \nu\pi + 2\pi A \cos u_{j + 1} .& (\+b)\cr}$$
Combining (\foo) and (\+) we obtain the following difference
equations which give the $(j + 1)$th orbit in terms of
the $j$th orbit,
$$\vjust{\eqaligntwo{u &= \t - \r ,&  v &= \t + \r ,& (\en{de}{13}a)\cr\eqskp
\t &= \half(v + u) ,&  \r &= \half(v - u) ,& (\+b)\cr}
\eqskip
\eqalignno{u_{j + 1} - u_j &= 2\pi\d - 2\pi A \cos v_j ,& (\+c)\cr\eqskp
v_{j + 1} - v_j &= 2\pi\d + 2\pi A \cos u_{j+1} ,& (\+d)\cr}}$$
\xdef\de{\eqprefix {13}}%
where
$$A = {\a\nu\over r} \abs{\Hp(r)} , \eqn{\en{Adef}{14}}$$
\xdef\Adef{\eqprefix {14}}%
$$\d \eqv \nu - n , \eqn{\en{ddef}{15}}$$
\xdef\ddef{\eqprefix {15}}%
$\H$ is the Hankel function of the first kind and of
order $\nu$, $\abs{\d} \le \half$, and $n$ is an integer.  In
writing (\Adef) we have made use of the expansion of the
Bessel functions for $r > \nu + \nuth$ [Ref.\ 3, Eq.\ (9.3.3)].  Writing
$A$ in this way is mainly a notational convenience, although it
comes out naturally in this form when we use the whole cyclotron
orbit as the unperturbed orbit.\ref{4}  We treat $A$ as being a
parameter (independent of $\r$) in (\de).  We
shall define $T$ as the mapping which takes a point to its
iterate:
$$T(\t_j , \r_j) = (\t_{j+1}, \r_{j+1}) . \eqn{\en{Tdef}{16}}$$
The limits of validity of (\de) are found to be (see Appendix A)
$$\nu \grgr 1,\qquad r-\nu \grgr \nuth,\qquad
A \lsls (r^2-\nu^2)^{3/2}/r^2 . \eqn{\en{foo}{17}}$$
\xdef\foo{\eqprefix {17}}%

Note that the problem now just depends on two parameters,
$\d$ and $A$.  This was recognized in I\null, but was only established
analytically in the case $r \grgr \nu$, which is
much more restrictive than (\+).

The difference equations, (\de), may be iterated numerically and compared with
the solution to the Lorentz force law, (\lfl).  Such
a comparison is made in Fig.\ 2, where Fig.\ 2(a) is taken from I
and Fig.\ 2(b) is obtained from the difference equations.  We
see that there is excellent
agreement, so that the difference equations indeed
provide an accurate approximation to the Lorentz force law.

We note that the kicks received by the ions happen at the
Landau resonance points (Fig.\ 1), and so it is interesting
to compare the ion motion under the conditions described
by (\+) with Landau damping in the absence of a magnetic
field.  There are two important differences.  Because
the magnetic field sweeps the vector, $\vec{v}_\perp$,
through all angles, the resonant particles in our
case are those for which $r > \nu$ or $v_\perp > \omega/k_\perp$.  These
particles are much more numerous than those satisfying the
normal Landau resonance condition $\.y = \nu$ or
$v_y = \omega/k_\perp$.  The second difference is that the
kick that the particle receives, $B$, is a function of the
magnetic field, since the magnetic field causes the particle to spend
only a short time in resonance.  In the linear limit of normal Landau
damping the resonant particles remain in resonance forever.

\section{III}{Simple Properties of the Difference Equations}

The difference equations, (\de), describe the mapping
of the $(\t, \r)$ plane onto itself.  It models
the nonlinear coupling of two harmonic oscillators, the parameter,
$\d$, giving the relation between the unperturbed
frequencies and $A$ giving the strength of the coupling.  The
mapping is area-preserving, being derivable from the generating
function,
$$\eqalignno{F(u_{j+1}, v_j) &= u_{j+1}v_j + 2\pi\d(u_{j+1}-v_j)
+2\pi A(\sin u_{j+1} + \sin v_j) . &(\e)\cr}$$
[Equations (\de c) and (\de d) are given by $u_j = \pd F/\pd v_j$
and $v_{j+1} = \pd F/\pd u_{j+1}$, respectively.]  Thus, (\de)
describes a Hamiltonian system.

The $(\t, \r)$ plane is periodic (period $2\pi$) in both
the $\t$ and $\r$ directions and so is topologically equivalent
to a $2\pi \times 2\pi$ torus.  The periodicity in $\r$ is a consequence
of taking $A$ to be independent of $\r$.  This, in turn, will
allow a very simple determination of the diffusion coefficient
(see Secs.\ VII and VIII).  The periodicity in
$u$ and $v$ allows a stronger statement to be made, viz.,
the transformation,
$$\t \to \t + \pi ,\qquad \r \to \r \pm \pi , \eqn{\e}$$
leaves (\de) invariant.  This means it is sufficient to examine a range
of $2\pi$ in $\t$ and $\pi$ in $\r$.  Other transformations
that leave the difference equations invariant are:
$$\eqalignthree{\d &\to \d +1; &&&&& (\e)\cr\eqskp
\d &\to -\d,& \t&\to \pi-\t,& \r &\to -\r;& (\e)\cr\eqskp
A&\to -A,& \t&\to \t-\pi ;&&& (\e)\cr\eqskp
j&\to -j,& \t&\to \pi - \t .&&& (\en{timinv}{23})\cr}$$
\xdef\timinv{\eqprefix {23}}%
The first three of these allow a further restriction
of the problem to $0 \le \d \le \half$ and $A \ge 0$.  The
last transformation shows the invariance of (\de) to time
reversal.

\section{IV}{Small Amplitude Solution}

In this and the next section we examine some aspects of the
transition from coherent to stochastic behavior,
beginning with the analysis for small $A$.

It is possible to write down the Hamiltonian for the difference
equations, (A13).  For small $A$, canonical transformations
similar to those in I may be performed, which allow a solution
to the problem.  Here we take a different approach which
achieves the same results in a more direct manner.

In the limit, $A = 0$, (\de) describes two uncoupled harmonic
oscillators.  If $\d$ is a rational number, $s/p$, then all
the points in the $(\t, \r)$ plane are $p$th-order fixed points.  If
we now let $A$ be small, we expect $\t_p - \t_0$ and
$\r_p - \r_0$ to be small also.  In that case the difference equations
may be approximated be differential equations for $\.\t$ and
$\.\r$.  (It might seem that we are going round in circles
here.  However, these differential equations will be much simpler
than the exact equations of motion,
since they describe motion with only one degree of freedom.)  If we
can find the Hamiltonian, $h$, such that $\.\t = \pd h/\pd \r$
and $\.\r = - \pd h/\pd \t$, then we have found an additional constant
of the motion, $h$.  Lines of constant $h(\t, \r)$ give the
trajectories of ions in the $(\t, \r)$ plane.

We begin with the case, $\d = 0$ (i.e., a cyclotron
harmonic).  Then (\de) becomes
$$\eqalignno{u_1 - u_0 &= - 2\pi A \cos v_0, &(\e a)\cr\eqskp
v_1-v_0 &= 2\pi A \cos u_1 = 2\pi A \cos u_0 + O(A^2), &(\+b)\cr}$$
so that
$$\eqalignno{\t_1 -\t_0 &= 2\pi A \sin \t_0 \sin\r_0 + O(A^2), &(\e a)\cr\eqskp
\r_1 -\r_0 &= 2\pi A \cos \t_0 \cos\r_0 + O(A^2). &(\+b)\cr}$$
These equations are Euler's approximation to the solution of
the differential equations,
$$\eqalignno{\.\t &= A\sin\t\sin\r, &(\e a)\cr\eqskp
\.\r &= A \cos\t\cos\r. &(\+b)\cr}$$
[We define $\t_j = \t(t = 2\pi j)$, etc.]  These, in turn, are Hamilton's
equations for the Hamiltonian,
$$h = -A\cos\r\sin\t. \eqn{\en{hzero}{27}}$$
\xdef\hzero{\eqprefix {27}}%
The trajectories of the ion in $(\t, \r)$ space,
given by this Hamiltonian, are shown in Fig.\ 3(a).  This result is the
same as that obtained by Timofeev\ref{5} using the full Hamiltonian
for the ion.  Note that all particles exchange energy with the
wave, but that the amount of energy exchanged is bounded in time.

It is instructive to compare this result with standard linear
theory.  As we noted, our approach is nonlinear because we repeatedly
correct the unperturbed orbits.  If this is not done (almost)
all the orbits are secular with $\r_N - \r_0 \sim AN$.  These orbits
would appear as straight lines in the $(\t, \r)$ plane which
are tangent to the curves drawn in Fig.\ 3(a) at the initial
positions of the ions.  Because all the orbits are secular, the linear
dielectric function is divergent at a cyclotron harmonic (for $k_\para
= 0$).  Linear theory is valid for $t \lsapprox \tau$ where $\tau$
is the inverse frequency of motion around the islands in
Fig.\ 3(a).  From (\+) we find $\tau = A^{-1}$ (near the center
of the islands).

For $\d = \half$ (i.e., $\omega$ at a half harmonic of
$\Omega_i$) we find
$$\eqalignno{u_2 - u_0 &= 2\pi+4\pi^2 A^2 \sin v_0\cos u_0 +O(A^3), &
(\e a)\cr\eqskp
v_2 - v_0 &= 2\pi-4\pi^2 A^2 \cos v_0\sin u_0 +O(A^3). &(\+b)\cr}$$
Ignoring the shift of $u$ and $v$ by $2\pi$ the
differential equations for $\t$ and $\r$ are
$$\eqalignno{\.\t &= \pi A^2\sin 2\r , &(\e a)\cr\eqskp
\.\r &= - \pi A^2\sin 2\t , &(\+b)\cr}$$
which are derivable from the Hamiltonian,
$$h = -\pi A^2 \cos(\t+\r)\cos(\t-\r). \eqn{\e}$$
The trajectories for this case are shown in Fig.\ 3(b).

The motion is very similar to the case, $\d = 0$, Fig.\ 3(a).  The
most important difference is the extra factor of $A$ in (\+) as compared
with (\hzero) so that the exchange of energy with the wave is much slower
at a half harmonic.  Indeed, to order $A$ there is no
exchange of energy since the phase of the ion relative to the
wave increases by about $\pi$ each cyclotron period, and so
the kicks received by the ion on successive cyclotron orbits
nearly cancel.  Thus, there is no contribution to the linear
dielectric function from the half harmonics.  We expect the $O(A^2)$
change in the speeds of the particles in this case to lead to
additional nonlinear terms in the dielectric function which would
cause a nonlinear damping of the waves due to half-harmonic
resonances.

In the general case where $\d = s/p \ne 0, \half$, we have
$$\eqalignno{\t_p - \t_0 &= 2\pi s + p\pi^2 A^2
{\cos(2\r - \d\pi) \over \sin(\pi\d)} + O(A^3), &(\e a)\cr\eqskp
\r_p - \r_0 &= O(A^3).& (\+b)\cr}$$
These expressions were explicitly checked only for $\d = \fract{1}{12}$,
$\fract{1}{8}$, $\fract{1}{6}$, $\fract{1}{4}$, $\fract{1}{3}$, $\fract{3}{8}$,
$\fract{5}{12}$.  The algebra required to do the iterations and
expansions of the difference equations is quite onerous, and so was
performed using the algebraic manipulation system,
M{\sc ACSYMA}.\ref{6}  We note that the expression for
$\.\t$ given by dividing (\+a) by $2\pi p$ agrees with the
expression for the frequencies given by Eq.\ (43) of I\null.  [This
result in I was derived using canonical transformations on the
the Hamiltonian, (A1), and required considerably more
effort than (\+).]  The trajectories in this case are
$\r = \hjust{const.}$ [Fig.\ 3(c)].  In this case there is no net exchange
of energy between the particles and the wave.

\section{V}{The Stochastic Transition}

For finite $A$ the motion becomes much more complicated.  Here
we briefly discuss what happens.  The exposition is intended to
complement that given in I\null,
although there is some overlap.  General
reviews of the transition to stochasticity may be found in Refs.\ 7--9.

We begin by noting the topological difference between the cases
$\d = 0$, $\half$ and $\d \ne 0$, $\half$ (Fig.\ 3).  This
leads to differences in the way that stochasticity develops, and
because of this we define the stochasticity threshold
differently in the two cases.

With $\d$ not close to 0 or $\half$, higher-order terms in the
difference equations, i.e., the $O(A^3)$ terms in (\+),
cause islands to appear.  The location
of the islands may be derived by requiring that $(\t_p - \t_0)
/(2\pi p)$, (\+a), be rational.  Where these islands
overlap the motion becomes stochastic [Fig.\ 2(b)].  However,
the KAM (Kolmogorov--Arnold--Moser)
theorem\ref{10} ensures that for sufficiently small $A$
some trajectories exist which span $\t$ [again see Fig.\ 2(b)].  [These
are KAM ``surfaces'' reflecting their dimensionality in the
original problem, although they only appear
as lines in the $(\t , \r)$ plane.]  These act as barriers,
preventing $\r$ for a given particle from
increasing or decreasing by more that about $\pi$ (since the
KAM surfaces repeat periodically in $\r$).  At some
value of $A$ the last KAM surface disappears, allowing unrestriced
motion in $\r$.  We define this value of $A$ to be the
stochasticity threshold, $A_s$.  In I the value of $A_s$ was
found to be about $\quarter$.  Recently accurate methods for
determining $A_s$ have been developed by Greene,\ref{11}
although applying these methods to the problem in hand presents
difficulties because the relevant resonances do not exist down to
$A = 0$.

For $\d$ equal to 0 or $\half$, we have, for small $A$, all
of phase space covered by islands [Fig.\ 3(a) and (b)].  When
$\d$ is not equal to 0 or $\half$ but is close to these values,
Fig.\ 3(c) applies only for very small amplitudes.  If $A > \abs{\d}$
or $A > (\abs{1-2\d}/\pi)^{1/2}$, first- and second-order islands
appear.  At larger amplitudes these islands can cover nearly
all of phase space giving a picture similar to Fig.\ 3(a)
and (b).  The effect of finite $A$ in these cases is to produce
chains of islands within the main islands.  These overlap close
to the separatrix causing a layer around the separatrix to become
stochastic.  For $\d = 0$ the thickness of this layer is found\ref{12}
to be on the order of
$$\exp[-\pi/(2A)] \eqn{\e}$$
for small $A$;  a similar exponential dependence
is expected for $\d = \half$.  Thus, in contrast with the previous
case it is possible for ions to be heated even at low
amplitudes.  However, for small $A$,  the thickness of the
stochastic layer, (\+), is extremely small.  This means firstly
that only a tiny fraction of particles will be in the
stochastic region and secondly that those that are will be
heated very slowly.  Therefore we define the stochasticity
threshold in this case as the value of $A$, $A_s$, for which the
stochastic layer occupies a substantial fraction of phase space.  Note
that this does not define a precise value of $A_s$.  The definition
is useful however since (\+) is a strong function of $A$.  In
I it was found that $A_s$ had approximately
the same value as for $\d \ne 0$, $\half$, namely $\quarter$.

The stochasticity threshold, $A_s = \quarter$, has a simple
physical interpretation.  At this value of $A$, the kick received
by the particle on one transit through
wave-particle resonance (Fig.\ 1) is sufficient to change the
phase that the particle sees when next in wave-particle resonance
by $\pi/2$ ($\null = 2\pi A_s$).  This explanation, which
is valid for $r > \nu$, complements the explanation in terms
of trapping for the case $r \approx \nu$, which was given in I\null.

We end this section by looking at the eigencurves of the
mapping.  These are curves in the $(\t, \r)$ plane which map
into themselves on applying the mapping, $T$.  A study of the
eigencurves provides another view of the stochastic transition.  Again
the reader is referred to Refs.\ 7--9 for a fuller discussion of this.  We
consider only the case $\d = 0$.  (Because of the additional
symmetries in this case, the eigencurves are easier to
generate.  Similar behavior is observed for other values of $\d$.)  When
$A$ is infinitesimal the eigencurves are given by Fig.\ 3(a);  and
Fig.\ 4 shows them for finite values of $A$.  For finite but small
$A$, most of the eigencurves remain closed;  these may be generated
by iterating the mapping of a single point many times.  However, the
eigencurves emanating from the hyperbolic fixed points no longer
meet up.  They are now infinitely long open curves.  They are
generated as follows:  Close to the hyperbolic fixed point, the
mapping may be linearized.  The eigencurves of the
linearized mapping are hyperbol\ae.  The
asymptotes of the hyperbol\ae\ are eigencurves on which a point either
moves away from or towards the fixed point.  These eigencurves
are called the unstable and stable manifolds respectively.  Short
straight line segments, which connect a point with its image under $T$,
are chosen on these unstable and stable manifolds.  These
line segments are then repeatedly mapped forwards and backwards by
applying $T$ and $T^{-1}$;  and the union of the mapped segments
generates the open eigencurves.  In practice, only a few
(less than 10) iterations are required.  Figure 4(a) shows the eigencurves
for $A = 0.3$.  We see that the open curves intersect each
other at a finite angle, and start oscillating as they
approach the fixed point.  Because the mapping is
area-preserving these oscillations grow larger and larger
as the fixed point is approached.  These curves occupy a finite
area which may be shown to be stochastic.  As
illustration of this we show the iteration of a single point
starting close to the hyperbolic fixed point (Fig.\ 5).  We
see that it covers nearly all the area outside
the first-order islands.  All these points lie on
a single eigencurve.  The area occupied by the eigencurves
(i.e., the area of phase space that is stochastic) is proportional
to the angle at which the eigencurves first intersect
(approximately half way between the fixed points).  This then
gives another way, in addition to the method of
island overlap, of estimating the fraction of phase space
that is stochastic.  This angle again has the exponential dependence
on amplitude given in (\+).  In Fig.\ 4(b), $A$ has been increased
to 0.35.  Here we see the open eigencurves intersecting at
a larger angle.  In addition, the elliptic fixed points in
Fig.\ 4(a) have turned into hyperbolic fixed points
with reflection [from (B7) we find that
this happened at $A = \pi^{-1} \approx 0.318$].  Thus,
open eigencurves emanate from these points.  However, the angle
at which they intersect
is so small that the area they occupy is comparable to the
thickness of the lines used to draw the figure.  Eigencurves
corresponding to the original islands are still present.  However,
these have disappeared at $A = 0.45$, Fig.\ 4(c).  At this
amplitude the two sets of open eigencurves intersect each other.  This
may be seen by comparing this figure with
Fig.\ 4(d), where we have extended one of the
eigencurves emanating from the hyperbolic fixed point with reflection.

\section{VI}{The Krylov--Kolmogorov--Sinai Entropy}

Having established that the ion motion becomes stochastic for
$A > A_s$, we need to be able to describe the motion above the
stochasticity threshold.  One of the most important parameters in
this regard is the Krylov--Kolmogorov--Sinai (KS) entropy, $h$.  This
is a measure of the local instability of trajectories and is defined
as the average rate of divergence of neighboring
(infinitely close) trajectories.  Thus, after $N$ iterations the
distance between neighboring particles initially $d_0$ apart is
approximately $d_0 \exp(hN)$ (for $d_0$ small and $N$ large).

The KS entropy is important for two reasons.  Firstly, given a group
of particles which initially occupy a small region of phase
space, $\Delta\t\,\Delta\r$, we can estimate the number
of iterations for the phases, $\t$, of the particles to become
random to be
$$\log (2\pi/\Delta\t)/h .\eqn{\en{foo}{33}}$$
\xdef\foo{\eqprefix {33}}%
Secondly, the exponential divergence of trajectories leads to a
{\ital mixing} of phase space, which in turn results in the decay
of the correlation function, $C_k$ (Sec VII), and allows the
motion of the particle to be described by a diffusion equation.

In order to determine $h$ we define a reference trajectory,
$(u_j, v_j)$.  [It is more convenient to work in
$(u, v)$ space.]  We consider a neighboring trajectory, $(u_j + \d u_j,
v_j + \d v_j)$, where $\d u_j$ and $\d v_j$ are infinitesimals.  Then
$\d u_j$ and $\d v_j$ satisfy
$$\onecol{\d u_{j+1} \cr \d v_{j+1} \cr} = \mat{J}_j
\onecol{\d u_j \cr \d v_j \cr} , \eqn{\en{linmap}{34}}$$
where
$$\mat{J}_j \eqv \twocol{1 & 2\pi A \sin v_j \cr\caseskp
-2\pi A\sin u_{j+1} & 1- 4\pi^2 A^2 \sin u_{j+1} \sin v_j\cr} .
\eqn{\en{Jdef}{35}}$$
\xdef\Jdef{\eqprefix {35}}%
If we define $\mat{M}_N$ by
$$\mat{M}_N = \prod_{j = 0}^{N-1} \mat{J}_j , \eqn{\e}$$
then $h$ is given by
$$h = \Lim_{N \to \inf}(\log \abs{\Lambda_N}/N) , \eqn{\e}$$
where $\Lambda_N$ is the largest (in magnitude) eigenvalue of
$\mat{M}_N$.  This definition is equivalent to\ref{9}
$$h = \ave{\log(\l_{j+1}/\l_j)} , \eqn{\e}$$
where $\l_j$ is the length of the vector $(\d u_j, \d v_j)$
and the average is taken over a particle orbit.

Figure 6 shows $\log\abs{\Lambda_N}/N$ as a function of $n$
for various values of $A$.  We see that it does have a limiting value
as $N \to \inf$ and so $h$ is well defined.  Figure
7 shows $h$ as a function of $A$ for $\d = 0.23$.  For
$A$ small, $h$ is close to zero, because when the motion is coherent neighboring
trajectories diverge linearly, rather than exponentially.  When
$A$ is still below $A_s$, $h$ becomes finite, since the reference
trajectory was chosen in the stochastic part of
phase space (Fig.\ 2).  As $A$ passes through the stochasticity
threshold there is a slight drop in $h$.  Finally, for $A
\grgr A_s$, $h$ is on the order of $\log A$.

We may derive the expression for $h$ in the limit $A \grgr A_s$.  If
$\lambda_j$ and $\mu_j$ are the larger and smaller (in magnitude)
eigenvalues of $\mat{J}_j$, then for $A \grgr A_s$
we have
$$\lambda_j \approx -4\pi^2 A^2 \sin u_{j+1} \sin v_j ,\qquad
\mu_j = 1/\lambda_j . \eqn{\en{leq}{39}}$$
\xdef\leq{\eqprefix {39}}%
Furthermore, the eigenvectors corresponding to these
eigenvalues are ap\-prox\-i\-mate\-ly $\^{\vec{v}}$ and
$\^{\vec{u}}$ respectively.  Since the eigenvectors for different
$\mat{J}_j$ are nearly parallel to one another, the eigenvalues
of the product of the $\mat{J}_j$ are
approximately the product of the eigenvalues of the $\mat{J}_j$;  i.e.,
$$\Lambda_N \approx \prod_{j = 0}^{N-1} \lambda_j . \eqn{\e}$$
Substituting (\leq) into (\+) we have
$$h = \ave{\log\abs{\lambda_j}} = 2\log (\pi A) + 
\ave{\log\abs{4\sin u_{j+1} \sin v_j}}, \eqn{\e}$$
where the average is taken over a particle orbit.  Finally,
we note that for $A \grgr A_s$, a particle's orbit wanders
over most of phase space, spending equal amounts of time
in equal areas of phase space (i.e., the particle's
orbit is approximately ergodic).  Thus, the average along the orbit
may be replaced by a phase space average.  The last term in (\+)
becomes
$$2 (2\pi)^{-1} \int_0^{2\pi}\log\abs{2 \sin\psi}\,d\psi = 0 \eqn{\e}$$
[see Ref.\ 3, Eq.\ (4.3.145)].  Thus, for $A \grgr A_s$, we have
$$h \approx 2 \log(\pi A) . \eqn{\en{hres}{43}}$$
\xdef\hres{\eqprefix {43}}%
This is shown as a dashed line in Fig.\ 7.

\section{VII}{The Correlation Function}

We saw in the last sections that a group of
particles initially close together in phase space
separate exponentially.  This continues until a time
on the order of $h^{-1}$, (\foo).  After this
time the phase of the particles may take on any
values.  We wish to ascertain the behavior of
the particles for longer times.  More
precisely, if we consider an ensemble of particle
trajectories, $(\t_j, \r_j)$, then what are the
moments, $\ave{\r_N - \r_0}$, $\mtwo$, for large $N$?  (The
brackets denote ensemble averages.)

Firstly, let us consider the average force on the
ensemble,
$$f \eqv {\ave{\r_N - \r_0}\over N} . \eqn{\e}$$
By shifting the origin of time by $N$ (which leaves
the difference equations invariant) we have
$$f = {\ave{\r_0 - \r_{-N}}\over N} , \eqn{\e a}$$
and by reflecting time using the symmetry,
(\timinv), we have
$$f = {\ave{\r_{-N} - \r_0}\over N}. \eqn{\+b}$$
From (\+a) and (\+b) we see that $f = 0$.

Turning to the second moment, we first define an
acceleration.
$$a_j = \r_{j+1}-\r_j . \eqn{\e}$$
(In defining $a_j$, we do {\ital not} treat
$\r$ as a periodic variable.)  Then we have
$$\mtwo = \sum_{i=0}^{N-1} \sum_{j=0}^{N-1} \ave{a_i a_j} .
\eqn{\en{moment}{47}}$$
\xdef\moment{\eqprefix {47}}%
We define a correlation function, $C$, by
$$C\left( {i+j\over 2}, i-j \right) = \ave{a_i a_j}. \eqn{\e}$$
$C$ has the following properties:
$$C(\l, k) = C(0, k) \eqv C_k , \eqn{\e a}$$
which follows from the invariance of the system to a shift in the
origin of time, and
$$C_k = C_{-k} , \eqn{\+b}$$
since $\ave{a_i a_j} = \ave{a_j a_i}$.  Because of (\+a) the
definition of $C_k$ may include an average over a single orbit,
as well as an ensemble average.  [We will usually use a subscript,
$N$, to denote the time (i.e., the iteration number), while
a subscript, $k$, will denote a time difference.]

Rewriting (\moment) in terms of $C_k$ and using
properties, (\+), we find
$$\mtwo = N C_0 + \sum_{k=1}^{N-1} 2(N-k) C_k . \eqn{\e}$$
For $N \to \inf$ we have
$$\mtwo = 2\D N, \eqn{\e}$$
where the diffusion coefficient, $\D$, is given by
$$\D = {1\over 2} C_0 + \sum_{k=1}^{\inf} C_k  \eqn{\en{Ddef}{52}}$$
\xdef\Ddef{\eqprefix {52}}%
and we have assumed that $C_k$ decays sufficiently rapidly so
that the sum in (\+) exists.

In order to find $C_k$ numerically we compute $M$ orbits of
length, $L$.  Then $C_k$ is approximated by
$$C_k = (L-k+1)^{-1} \sum_{\l=0}^{L-k} \ave{a_{\l}a_{\l+k}} ,
\eqn{\en{Ccomp}{53}}$$
\xdef\Ccomp{\eqprefix {53}}%
where the average is now over the $M$ orbits.

Figure 8 shows examples of the correlation function for
$\d = 0.23$ and various values $A$.  When $A < A_s$,
Fig.\ 8(a), $C_k$ is highly oscillatory, and decays slowly.  As
$A$ is increased above the stochasticity threshold, the
correlation time, $k_c$, rapidly decreases, Fig.\ 8(b), becoming
approximately 0 for $A \grapprox 1$, Fig.\ 8(c).

Note that in defining the ensemble of orbits we did not
specify what $\r_0$ was, so that different members
of the ensemble could have different values of $\r_0$.  Because
of the periodicity of the difference equations, (\de),
in $\r$, we know that $\r_0 \mod{2\pi}$ is sufficient
to determine $\r_N - \r_0$.  Thus, the ensemble average
provides an average of the initial conditions over
an interval of $2\pi$, which is of little
consequence.  Averaging along a given orbit (for $A > A_s$)
has the same effect.  Having performed this averaging,
$C_k$ is independent of $\r$.

The situation is much more complicated in the original
system described by the Lorentz force law, (\lfl).  In
that case the problem is not periodic in $r$, since
$A$ has a slow dependence on $r$, (\Adef).  Now,
we would be hiding this important dependence on $r$
if we did not restrict the starting positions of the orbits,
$r_0$.  Similarly, we could not perform the averaging along
the orbit.  Thus, $C_k$ becomes a function of $r$, and it is
more difficult to compute, since less averaging
can be done.  Finally, the diffusion coefficient would
be more difficult to define because, although $\ave{(r_N-r_0)^2}$
might be proportional to $N$ for some range of $N$, it certainly will not
be proportional to $N$ for large $N$, since
the particles will be coming into regions where
$A$ and $C_k$ are different.

These problems are all circumvented by using the difference
equations.  These allow a simple and rapid determination
of $C_k$ and $\D$.  The dependence of
the diffusion coefficient on $r$ will be recovered through
its dependence on $A$ (Sec.\ IX).

\section{VIII}{The Diffusion Coefficient}

We now turn to the diffusion coefficient for the ions.  We shall
use (\Ddef) to define and calculate $\D$.  In the limit
$A \grgr A_s$ we can obtain an estimate for $\D$ based
on the observation of the last section that the
correlation time is short in this limit,
viz., $C_k = 0$ for $\abs{k} > 0$ [see Fig.\ 8(c)].  In
this case we have $\D = \half C_0$, with $C_0 =
\ave{a_0^2}$, where the average is taken over an ensemble and
down a trajectory.  In the same spirit that we calculated
the KS entropy, $h$, in the limit $A \grgr A_s$ (Sec.\ VI),
we shall assume that the ion trajectories are ergodic
so that the ensemble and trajectory averages may be replaced by a phase
space average.  Thus, we obtain
$$\eqalignno{\D &= \half \ave{a_0^2} =
\fract{1}{8} \ave{[(v_1-v_0)-(u_1-u_0)]^2}\cr\eqskp
&= \half \pi^2 A^2 \ave{(\cos u_1 + \cos v_0)^2}\cr\eqskp
&= \pi^2 A^2 (2\pi)^{-1}\int_0^{2\pi} \cos^2 \psi\,d\psi\cr\eqskp
&=\half \pi^2 A^2 .&(\e)\cr\eqskp}$$

When $A$ is not large we expect the presence of islands
to obstruct the diffusion of ions leading to a value of
$\D$ below (\+).  [This is reflected in the correlation
function by the fact that for small $k$, $C_{k > 0} < 0$;
see Fig.\ 8(b).]  Finally, for $A < A_s$ diffusion is
stopped.  (Strictly speaking this only happens when $\d$ is
not close to 0 or $\half$, for then there are KAM surfaces
preventing the diffusion of ions.  However, as discussed in
Sec.\ V\null, the diffusion for $A<A_s$ for $\d$ close to 0
or $\half$ is extremely slow.)  We therefore expect $\D$
to have the form,
$$\D = \half \pi^2 A^2 g^2(A), \eqn{\en{Gdef}{55}}$$
\xdef\Gdef{\eqprefix {55}}%
where $g(A) \le 1$, $g(\abs{A} \to \inf) \to 1$, and
$g(\abs{A} < A_s) = 0$.

We test these ideas out by numerically finding $\D$,
and hence $g(A)$, for various values of $\d$.  We determine
$\D$ by first finding $C_k$ using (\Ccomp) and then computing
$$\D = \half C_0 + \sum_{k=1}^K q_k C_k, \eqn{\en{Dcomp}{56}a}$$
\xdef\Dcomp{\eqprefix {56}}%
where $q_k$ is a windowing function given by
$$q_k = \case{1,&\for 0 < k \le K^\prime,\cr\caseskp
{1\over 2} \left[1+\cos\left(\pi {k - K^\prime\over K+1-K^\prime}
\right)\right],&\for K^\prime < k \le K .\cr}\eqn{\+b}$$
The reason for introducing a windowing
function into (\Dcomp) is to suppress the
effect of the oscillations seen in Fig.\ 8(a).  Criteria for
choosing $K$ and $K^\prime$ are
$K-K^\prime \grgr 1$ for $A < A_s$ (in order to suppress
the oscillations in $C_k$),
$K^\prime \grgr k_c$ for $A > A_s$ (so as to include the
major contribution to $\D$), and
finally $K \lsls L$, where $L$ is the length
of the orbits used to calculate $C_k$, (\Ccomp).  This
last condition arises because, from (\Ccomp), the
number of samples entering into $C_k$ is
$M(L-k+1)$ which we wish to be large in order to suppress
the statistical fluctuations in $C_k$.

Note that if we had computed $\D$ directly by
measuring $(\r_L - \r_0)^2$ for $M$ orbits, there
would be only $M$ samples contributions to $\D$.  Thus, the error
in $\D$ would be $O(M^{-1/2})$.  When computing
$\D$ using $C_k$ we effectively include all possible shorter
orbits that make up the orbit $\r_0$ to $\r_L$.  There
are about $L/k_c$ independent orbits in an orbit of length,
$L$.  The error in $\D$ is then $O[(k_c/ML)^{1/2}]$
which can be much less than the error when
computing $\D$ directly.

Figure 9 shows $g(A)$ for three value of $\d$.  Although
$g$ has the general form discussed earlier, there
are some anomalies.  Instead of approaching unity
monotonically as $A \to \inf$, there is a tendency for
$g$ to oscillate about unity with a period of unity
in $A$.  This is particularly noticeable with $\d$ equal
to 0.11 and 0.47.  The oscillations are about $\pm 20 \%$ for
$1 \le A \le 2$ and about $\pm 10 \%$ for $20 \le A \le 22$.  Also
there is a very strong peak in $g(A)$ for
$\d = 0.47$ and $A = 0.55$.  Both these phenomena are due to the
existence of ``accelerator'' modes, or island systems
for which the ion, rather than returning to its original
island, returns to an island displaced upwards
or downwards in $\r$ by an integer multiple of $\pi$.

Some of these accelerator modes are studied in Appendix B.  From
(B4) with $m = 1$ and $n = 0$ we find that there exists a stable
accelerating fixed point for $\d = 0.47$
and $0.53 < A < 0.5971$. So for $A = 0.55$ there are
orbits in the island system for which $\r_N \sim \pm
\pi N$.  Although the orbits used to calculate $\D$ all lay outside
this island system, they could wander close to the island
and become temporarily trapped close to the island (for
up to several hundred iterations).  This results in a very slow
decay of the correlation function (Fig.\ 10), and an
anomalously high diffusion coefficient.

As $A$ is increased, a new set of accelerator modes appear
at values of $A$ equal to $p \pm \d$, where $p$ is an
integer.  The islands disappear by becoming unstable, shortly
afterwards.  This leads to the oscillatory
behavior in $\D$ seen in Fig.\ 9.  Similar oscillation have
been observed by Chirikov\ref{9} for the
so-called standard mapping.  He also associated these
oscillations with the presence of accelerator modes.

In our case there are probably no accelerator modes
present in the cases for which $g(A)$ is plotted in Fig.\ 9
(except the case previously discussed, i.e., $\d = 0.47$,
$A = 0.55$).  The oscillations may however arise
from the unstable fixed points that remain after the islands
turn unstable.  Such unstable fixed points may marginally
increase the correlation time and hence the diffusion
coeficient.  As $\d$ is increased these fixed points
become more unstable and the effect is diminished.  Choosing
$A$ in the small windows where the accelerator modes
exist would have given peaks in $g(A)$ similar to the one
at $\d = 0.47$ and $A = 0.55$.  We may argue that
the peaks in $g(A)$ that occur when the accelerator modes
are present are unphysical.  In Appendix B we establish
that the window in $A$ for which these modes
exist is quite narrow.  Now, as an ion is being accelerated
or decelerated by one of these modes, $\r$ and $r$,
(\vdef a), will
be changing.  Hence, after a while we must recognize
that $A$, which is a function of $r$, (\Adef), will
also change.  This will ``switch off'' the accelerator modes after only
a few iterations, and the correlation times will similarly
be limited.  The effect of the slow change of
$A$ (which was ignored in deriving the difference equations) will
be to remove the peaks in $g(A)$, leaving
only the gentle decaying oscillations in $g(A)$.

For these reasons $g(A)$ was fitted by the simplest
smooth curve that rea\-son\-a\-bly fits the data.  Figure 9
shows the functional form chosen, viz.,
$$g(A) = \max\left({A^2-A_s^2\over A^2} , 0\right),\eqn{\en{geq}{57}}$$
\xdef\geq{\eqprefix {57}}%
with $A_s = \quarter$.  This form of $g$ will be checked in
the next section by solving the resulting diffusion equation,
and comparing the results with the solution of the exact equations
of motion, (\lfl).

\section{IX}{Derivation of a Diffusion Equation}

With $A > A_s$ we expect the distribution, $f(\r, N)$,
of a group of ions after $N$ iterations to spread
according to a diffusion equation.  We will
briefly outline the arguments that lead to this result.

Suppose $m$ is greater than the correlation time, $k_c$,
and let us look at the results of the mapping only at times,
$N$, which are an integer multiple of $m$.  Then the
mapping, $T$, may be approximated by a Markovian
process, which is described by a transition probability,
$P(\r_1\relv\r_2, N)\, d\r_2$, which is the probability
the a particle is found in the range, $(\r_2, \r_2 + d\r_2)$, given
that it was at $\r_1$, $N$ iterations earlier.  $P$ must satisfy
$$P(\r_1\relv\r_2, N) = \int d\r\,P(\r_1\relv\r,k)
P(\r\relv\r_2,N-k) \eqn{\en{Smo}{58}}$$
\xdef\Smo{\eqprefix {58}}%
for all $k$ such that $k$ is a multiple of
$m$ and $0<k<N$.

For long times $f$ will be spread over several periods in $\r$.  In
that case we are not interested in the details of $f$
that occur over the period, $2\pi$, in $\r$.  We retain the
important physical effects when we average $f$ and $P$ over a period
in $\r$.  (Precisely this averaging is carried out in
computing $C_k$ and $\D$ in Secs.\ VII and VIII\null.)  In
that case $P$ only depends on the difference, $\r_2 - \r_1$,
and we have
$$P(\r_1\relv\r_2, N) = P(\r_2-\r_1, N). \eqn{\e}$$

The evolution of the distribution of ions, $f$, is
given by
$$f(\r, N) = \int d\r_1\,f(\r_1, N_1) P(\r-\r_1, N-N_1).
\eqn{\en{fdef}{60}}$$
\xdef\fdef{\eqprefix {60}}%
The moments of $P$,
$$a_n(N) \eqv \int d\r\, \r^n P(\r,N), \eqn{\e}$$
are the same as the moments, $\ave{(\r_n-\r_0)^n}$, studied
in Sec.\ VII\null.  Thus, we have
$$\eqalignno{a_1(N)/N &= 0, &(\e a)\cr\eqskp
a_2(N)/N &= 2\D, &(\+b)\cr}$$
for $N \ge m$.

Following the analysis of Chandrasekhar\ref{13}
and Wang and Uhlenbeck\ref{14} we may use (\Smo) to derive a
{\FP} equation for the evolution of $P$ for $N \grgr k_c$:
$${\pd P\over \pd N} = {\pd^2\over \pd \r^2} \D P. \eqn{\en{Peq}{63}}$$
\xdef\Peq{\eqprefix {63}}%
Substituting for $f$ from (\fdef) gives
$${\pd  f\over \pd N} = {\pd\over \pd \r} \D {\pd\over \pd \r} f,
\eqn{\en{diffeq}{64}}$$
\xdef\diffeq{\eqprefix {64}}%
where we have used the result that $\D$ is independent of $\r$
in order to write the diffusion equation in its more usual
form.  Incidentally, although we derived (\+) for $A > A_s$
it clearly gives satisfactory results for $A < A_s$, because
$g$ and $\D$ are then zero, and $f$ does not evolve.

We now undo the normalizations made earlier.  Writing
$N = t/2\pi$ and converting $\r$ to $r$ using
(\vdef a) we have
$${\pd  f\over \pd t} = {1\over r} {\pd\over \pd r} r {\D\over 2\pi}
\left({\pd \r\over \pd r}\right)^{-2} {\pd\over \pd r} f,
\eqn{\e}$$
where $\pd \r/\pd r = (r^2-\nu^2)^{1/2}/r = \abs{\Hp(r)/\H(r)}$
for $r > \nu + \nuth$.  In writing (\+) in this form we have
replaced the first derivative operator in (\diffeq) by the
cylindrical divergence operator, in order to ensure conservation
of particles.  (The normalization of $f$ is such that
$\int dv_\para \int dr\, 2\pi rf =1$.  We will suppress the
dependence of $f$ on the velocity parallel to $\vec{B}$,
$v_\para$, until we consider the effects of collisions in
Secs.\ X--XII\null.)  Note that $\D$ is now a function
of $r$ through $A$, i.e., $\D = \D[A(r)]$, so that we
must justify our commuting $\D$ with the second derivative
operator between (\Peq) and (\diffeq).  The argument for
having $\D$ where it is in (\+) is that in the
steady state $f$ should be a constant in the stochastic region.  This
is a result of the approximate ergodicity of the motion,
when described by the exact equations of motion, (\lfl),
and was observed in the simulation described in I\null.  We
shall presently compare the results obtained
by the diffusion equation with that simulation.  Rewriting
(\+) using (\Gdef) and (\Adef) we obtain
$${\pd  f\over \pd t} = {1\over r} {\pd\over \pd r} r D
{\pd\over \pd r} f, \eqn{\en{diffnorm}{66}}$$
\xdef\diffnorm{\eqprefix {66}}%
where
$$D(r) = {\D\over 2\pi} \left({\pd \r\over \pd r}\right)^{-2}
={\pi\over 4} \left({\a \nu\over r} \abs{\H(r)} g(A)\right)^2
\eqn{\en{dnorm}{67}a}$$
\xdef\dnorm{\eqprefix {67}}%
for $r \ge \nu$.  We extend the definition of $D$ to $r < \nu$
as follows:
$$D(r) = \case{D(\nu), &\for \nu-\sqrt{\a}\le r<\nu,\cr\caseskp
D(\nu)[r-(\nu-2\sqrt{\a})]^2/\a,
&\for \nu-2\sqrt{\a}\le r<\nu-\sqrt{\a},\cr\caseskp
0, &\for r<\nu-2\sqrt{\a}.\cr}\eqn{\+b}$$
When numerically computing $D$ we use the approximate forms for
$H$ and $H^\prime$,
$$\eqalignno{\abs{\H(r)} &\approx (2/\pi)^{1/2} (r^2 - \nu^2)^{-1/4},
&(\en{Hdef}{68}a)\cr\eqskp
\abs{\Hp(r)} &\approx (2/\pi)^{1/2} (r^2 - \nu^2)^{1/4}/r, &(\+b)\cr}$$
for $r \ge \nu + \nuth$ and
$$\abs{\H(r)}\approx\abs{\H[\nu+\nuth]}, \eqn{\+c}$$
for $\nu \le r<\nu + \nuth$ and similarly for $H^\prime$.  We
will show the typical form of $D$ in Sec.\ XI\null.

Because of the restrictions on $r$ in the derivation of the
difference equations (Sec.\ II), we must
justify (\dnorm) for $r < \nu + \nuth$.  In this range, the choice
of the form of $D$ is motivated by the
following considerations:  Firstly, in the region,
$\nu -\sqrt{\a} < r < \nu+\nuth$ trapping\ref{4} plays a dominant
r\A ole.  This trapping is similar to the
trapping of a particle by a wave in the absence of a magnetic field
and its effect is to cause a rapid mixing of $f$ in this region.  This
is modeled by (\dnorm) since $D$ is approximately constant and
has its maximum value in this region.  Secondly, we find that
particles are fed quite slowly into the
trapping region from below and this is given by  the form of
$D$ for $\nu-2\sqrt{\a} < r < \nu-\sqrt{\a}$.  Finally, the motion is
coherent for $r < \nu-2\sqrt{\a}$ in which case $D$ is 0.

In order to check (\diffnorm) and (\dnorm) we compare the results
obtained using these equations
with the results of the simulation described in
I\null.  In the latter case the orbits of 50 particles were
found by integrating the exact equations of motion, (\lfl).  Because
of the small number of particles involved the comparison
is most easily made if we solve (\diffnorm) using a
simple Monte Carlo method described in Appendix C.  It might
seem as though we are back-tracking here, by going back
to difference equations.  However, the Monte Carlo
method for (\diffnorm) differs from the original equations,
(\de), in two respects:  firstly, the randomness is explicitly
inserted; and, secondly, we allow for the variation of $D$
with $r$.  This latter aspect of the solution we
present here allows us to justify the placement of $D$
with respect to the derivative operators in (\diffnorm).  Figures
11 and 12 show the comparison.  In each case
we follow the motion of $N = 50$ particles with
initial perpendicular velocities, $r_0 = 23$, and with $\a = 20$,
$\nu = 30.23$.  The general features of these figures are
discussed in I\null.  For our purposes we see that there is
good agreement between the two simulations.  In particular,
note the agreement in the short time behavior, Fig.\ 11(c) and (d),
and in the averaged distribution function, Fig.\ 12.  These
two points respectively confirm the form taken for $D$ for
$r \lsapprox \nu$ in (\dnorm b) and the placement of $D$ with respect
to the derivative operators in (\diffnorm).  We conclude that
(\diffnorm) with $D$ given by (\dnorm) accurately
models the diffusion of ions in the presence
of a perpendicularly propagating electrostatic wave.

We may write $D$ in dimensional form by multiplying
(\dnorm) by $\Omega_i^3/k_\perp^2$ to give
$$D(v_\perp) = \case{{\pi\over 4} {(E_0/B_0)^2\omega^2\Omega_i\over
k_\perp^2 v_\perp^2} \abs{\H(r)}^2 g^2(A)\gamma,
&\for v_\perp \ge \vph,\cr\caseskp
D(\vph),  & \for \vph-\vtr\le v_{\perp}<\vph,\cr\caseskp
D(\vph)[v_\perp-(\vph-2\vtr)]^2/\vtr^2,
&\for \vph-2\vtr\le v_{\perp} < \vph-\vtr,\cr\caseskp
0,&\for v_{\perp} < \vph-2\vtr,\cr}\eqn{\en{Dun}{69}}$$
\xdef\Dun{\eqprefix {69}}%
where $g(A)$, $\nu$, and $r$ are given by (\geq),
(\Adef), and (\pdef), $\vph=\omega/k_\perp$, and
$\vtr$ is the trapping velocity,
$$\vtr = (q_i E_0/m_i k_\perp)^{1/2}.\eqn{\en{vtreq}{70}}$$
In (\Dun) we have included an additional factor, $\gamma$,
which is defined as the fraction of time that the ion orbits
spend in the region of lower hybrid waves.  This reflects
the fact that the ion can only diffuse a fraction, $\gamma$,
of the time.  In order to include the spatial variation of the lower
hybrid waves in such a simple way, two conditions must
be satisfied.  The ions must spend many cyclotron periods in the lower
hybrid waves, so that the difference equations are applicable,
and the motion is described by (\diffeq).  The opposite
limit, where the ion spends a small fraction
of a cyclotron period in the wave, has been considered by
Lazarro\ref{15} in the context of heating a fast beam
of ions.  Secondly, we must be able
to model the lower hybrid wave as a product of a single
$\vec{k}$ component and a square-wave envelope.  The first condition
is usually satisfied in cases of practical interest and, because
the lower hybrid waves travel in well-defined rays,\ref{16}
the second condition is also normally satisfied.  In cases where the
envelope of the lower hybrid waves is not a square wave,
the definition of $D$, (\Dun), may be replaced
by the average of (\Dun), without the factor, $\gamma$,
over the ion's orbit.  For a
circulating ion in a tokamak which covers a magnetic
surface ergodically, $\gamma$ is approximately the ratio of the area
of the intersection of the lower hybrid ray with the magnetic
surface to the total area of the magnetic surface.  This
ignores the finite extent of the lower hybrid wave in the
perpendicular direction.  When this is included, $\gamma$ is reduced
since the cyclotron orbit must be completely in the lower hybrid
ray for the difference equations, (\de), to hold.  The diffusion
equation for the ions written in unnormalized form becomes
$${\pd f\over \pd t} = {1\over v_\perp} {\pd\over \pd v_\perp}
v_\perp D {\pd\over \pd v_\perp} f. \eqn{\en{diffun}{71}}$$
\xdef\diffun{\eqprefix {71}}%
In the next sections we study the properties
of this equation when we add a {\FP} collision term.

\section{X}{Inclusion of Collisions}

We have seen that the ions can irreversibly exchange energy with the
wave.  However, because $\omega/k_\perp$ is usually
several times the ion thermal speed, only the tail
ions are affected.  Bulk heating only takes place when these ions
collide with the background.  We investigate this process by including
a {\FP} collision term in (\diffun) to give
$${\pd f\over \pd t} = {1\over v_\perp} {\pd\over \pd v_\perp}
v_\perp D {\pd\over \pd v_\perp} f + \left({\pd f\over \pd t}\right)_c.
\eqn{\en{twod}{72}a}$$
\xdef\twod{\eqprefix {72}}%
Since only a relatively few tail particles are affected by
the wave we can neglect the tail-tail collisions when
determining $(\pd f/\pd t)_c$.  We can thus linearize this
term by assuming the the background distributions of
ions and electrons are Maxwellians.  We then obtain\ref{17}
$$(\pd f/\pd t)_c = -\suml_\beta \del\dot\vec{J}^{i/\beta}, \eqn{\+b}$$
where $\beta$ is a species label ($i$ or $e$),
$$\eqalignno{-\vec{J}^{\a/\beta}
&= {m_\a\over m_\a+m_\beta}\nuab_s\vec{v}f_\a
+{1\over 4} \nuab_\perp v^2 \del_{\subvec{v}}f_\a\cr\eqskp
&\qquad\null+\left({1\over 2}\nuab_\para-{1\over 4}\nuab_\perp\right)
\vec{vv}\dot\del_{\subvec{v}}f_\a,& (\+c)\cr}$$
and expressions for the collision frequencies, $\nu$,
can be found in Ref.\ 18.

We next simplify (\twod) by reducing it to a problem in only
one velocity dimension, $v_\perp$, by integrating (\twod a)
over $v_\para$.  The procedure we follow for doing this is very
similar to that employed by Fisch\ref{19, 20} to derive
a {\Oned} {\FP} operator in $v_\para$ in order to study
rf-driven currents.  When performing the integral over $v_\para$
the term involving $J_\para^{\a/\beta}$ drops out.  We
then need only to evaluate $-\int dv_\para\,
J_\perp^{i/\beta}$.  In order to do this we make three assumptions:  Firstly,
we assume that
$f$ ($= f_i$) has the form,
$$f(v_\perp,v_\para) = F(v_\perp) (2\pi T_i/m_i)^{-1/2}
\exp(-\half m_i v_\para^2/T_i); \eqn{\en{ffact}{73}}$$
\xdef\ffact{\eqprefix {73}}%
i.e., the dependence of $f$ on $v_\para$ is that of a
Maxwellian.  Secondly, we assume that $v_\perp^2 \grgr T_i/m_i$.  This
allows us to replace $v$ by $v_\perp$ in the expressions for the
collisions frequencies, so that they drop out of the
integral.  Lastly, we assume that the background temperatures are all
equal so that $T_\beta = T_i$.  (Without this assumption additional
sources and sinks of energy are required to maintain
an equilibrium.)  Then we have
$$-\int dv_\para\,J_\perp^{i/\beta} = C_\beta \,\pd F/\pd v_\perp
+ (m_i v_\perp/T_i)C_\beta F + O[T_i/(m_i v_\perp^2)], \eqn{\e}$$
where 
$$C_\beta = \half \nu_\para^{i/\beta} v_\perp^2 +
\quarter\nu_\perp^{i/\beta} T_i/m_i, \eqn{\e}$$
and we have used\ref{18}
$${m_\a\over m_\a+m_\beta}\nuab_s={m_\a v^2\over T_\beta}
{1\over 2} \nuab_\para . \eqn{\e}$$
Substituting these results into the integral of (\twod a) over
$v_\para$ gives
$${\pd F\over \pd t} = {1\over v_\perp} {\pd\over \pd v_\perp}
v_\perp \left[ D {\pd F\over \pd v_\perp} +
C\left({\pd F\over \pd v_\perp}  +
{v_\perp\over \vti^2}F\right)\right],
\eqn{\en{oned}{77}}$$
\xdef\oned{\eqprefix {77}}%
where $C = \sum_\beta C_\beta$ and $\vti^2 = T_i/m_i$.  If $D=0$,
we recover a Maxwellian for $F$.  Thus, we shall use (\oned) to solve
for $F$ for all $v_\perp$ even though it was derived only in the
high velocity limit.  We expect that this will not entail much
additional error as long as $D$ is finite only where $(v_\perp
/\vti)^2$ is large.

\section{XI}{Numerical Solution of Fokker--Planck Equation}

A number of approximations need to be made in the derivation
of the {\Oned} {\FP} equation, (\oned).  These
leave the accuracy of (\oned) open to question.  In particular
the factorization of $f$ by (\ffact) is not easy to rigourously
justify.  In this section we compare the numerical solutions of
the {\Oned} and {\Twod} {\FP} equations, (\oned) and (\twod).  A
similar, but more extensive, comparison has
been carried out\ref{21} for the parallel
{\Oned} {\FP} equation  for the electrons with a parallel
quasilinear diffusion term, derived by Fisch.\ref{19, 20}  For
reasons that we will discuss we expect better agreement
in the present case.  Therefore, we content ourselves with presenting here
the solution for one particular case.

Consider a hydrogen plasma with ion density, $n_0 = 10^{20}
\ \unit{m}^{-3}$, magnetic field, $B_0 = 4\ \unit{T}$, and temperature,
$T_e = T_i = 2 \ \unit{keV}$.  Let the lower hybrid
wave be described by $\omega/(k_\perp \vti) = 3.5$,
$\omega/\omega_{lh} = 1.3$, and $E_0 = 10^6 \ \unit{V}/\unit{m}$.  The
frequency of such a wave is $2.13 \ \unit{GHz}$.  Its parallel
wavenumber may be found by solving the dispersion
relation $K_\para k_\para^2 + K_\perp k_\perp^2
-3(\omega_{pi}^2\vti^2/\omega^4)k_\perp^4 = 0$, where
$K_\para$ and $K_\perp$ are elements of the
cold plasma dielectric tensor.  This gives $\omega/(k_\para v_{te})
= 5.25$ and a parallel index of refraction, $n_\para = 3.06$.  This
wave represents a typical Fourier component of
a lower hybrid wave near the center of the tokamak, but before it
has reached the point of thermal wave conversion.  Note that,
because of the high value of $\omega/(k_\para v_{te})$, we
expect electron Landau damping to be negligible.  The
normalized parameters are then found from (\pdef) to be
$\nu = 35$, $\a = 5.71$.  We take the extent of the lower
hybrid ray in the parallel direction to be $\Delta z =
0.4 \ \unit{m}$, the major radius of the tokamak to be
$R = 2 \ \unit{m}$, and the angular extent of the wave
in the poloidal direction to be $\Delta \t = \pi/4$.  The
geometrical factor, $\gamma$, appearing in (\Dun) is given by
$$\gamma \approx {\Delta z\over 2\pi R} {\Delta \t\over 2\pi}
={1\over 251}.\eqn{\e}$$
[As discussed in Sec. IX, this value of $\gamma$ should be slightly
reduced to account for the finite perpendicular extent of the
waves.  However, (\+) is sufficiently accurate for illustrative
purposes.]  Substituting these values into $D$, (\Dun),
at $v_\perp = \omega/k_\perp$
gives
$$D(\omega/k_\perp) = 1.5\, \vti^2 \nzero, \eqn{\en{dval}{79}}$$
\xdef\dval{\eqprefix {79}}%
where we have written $D$ in terms of a collisional
diffusion coefficient $\vti^2\nzero$ and $\nzero$ is the ion
collision frequency for thermal particles
$$\nzero = {\lambda_{ii} \omega_{pi}^4\over
4\pi n_0 \vti^3}.\eqn{\en{nudef}{80}}$$
\xdef\nudef{\eqprefix {80}}%
We take the Coulomb logarithm to be
$\lambda_{ii} = \lambda_{ie} = 15$.  With our parameters
$\nzero = 4.31\times 10^3 \ \unit{s}^{-1}$.  The full shape of
$D$ is given in Fig.\ 13.  This is used in the numerical
integrations of the {\FP} equations.

We now solve the {\Oned} and {\Twod} {\FP}
equations, (\oned) and (\twod), numerically.  At
$t = 0$ both $f$ and $F$ are taken to be Maxwellians with
temperature, $T_i$.  The diffusion coefficient is turned
on with a linear ramp function between $t = 0$
and $t = (\nzero)^{-1}$ and is kept constant thereafter.

Figure 14 shows the steady-state distributions.  [In plotting
the perpendicular distribution function in the {\Twod} case we
have defined
$$F(v_\perp) = \int dv_\para\, f(v_\perp, v_\para). \eqn{\e}$$
Note that this is consistent with (\ffact).]  Although
the dependence of $f$ on $v_\para$ is not that of
a Maxwellian, the two results for $F(v_\perp)$ are in
close agreement.  The major difference is that the tail of
the distribution function is characterized by a temperature
of $T_i$ in the {\Oned} case, but by about
$1.16\, T_i$ in the {\Twod} case.

The power dissipated by the wave is defined by
$$P_d = - 2\pi n_0 m_i\int v_\perp^2 D \,\pd F/\pd v_\perp\, dv_\perp,
\eqn{\en{pddef}{82}}$$
\xdef\pddef{\eqprefix {82}}%
while the power lost by the test distribution
to the background distributions is
$$P_c = n_0 m_i \int \suml_\beta \vec{v}\dot\vec{J}^{i/\beta}\, d\vec{v}
\eqn{\e}$$
in the {\Twod} case and
$$P_c = 2\pi n_0 m_i \int v_\perp^2 C[\,\pd F/\pd v_\perp
+ (v_\perp/\vti^2)F]\,dv_\perp \eqn{\en{pcdef}{84}}$$
in the {\Oned} case.  These are plotted as a function of time in
Fig.\ 15.  The most important quantity we wish to know is
the steady-state power dissipation, $P$.  As $t \to \inf$,
we have $P = P_d = P_c$.  We see that there is close agreement
between the {\Oned} and {\Twod} values for $P$.  (They differ
by about 3\%.)  However, the solution to the {\Oned}
equation reaches a steady state about 2 times faster than
that of the {\Twod} equation.  From Fig.\ 15(a)
we see that $P_d$ nearly reaches its steady-state value at
$t = \tau_d \approx 6/\nzero \approx 1.4\ \unit{ms}$,
whereas $P_c$ (and $f$ also) reaches a steady state on a time
scale, $\tau_c$, which is about $60/\nzero$ or
$14\ \unit{ms}$.

The agreement between the one-
and {\Twod} treatments in the power dissipated is much
better than observed in the case of the parallel
{\Oned} {\FP} equation for the electrons.\ref{21}  This
is so because the pitch-angle scattering terms, which are
the principle reason that $f$ deviates from the form assumed
in (\ffact) are approximately 4 times less important in the
case considered here.  Two effects contribute to this
factor of 4:  In our case electrons do not cause
appreciable pitch-angle scattering; whereas, in the
other case, ions and electrons contribute equally.  The other
effect is geometrical in origin and arises because
the perpendicular plane has two degrees of freedom, as
opposed to the parallel direction, which has only one.

We conclude that the {\Oned} {\FP} equation, (\oned),
gives accurate results for the steady state distribution and
for the power dissipated in the steady state, $P$.  The
agreement on the time scales is worse, although we expect the
{\Oned} equations to give qualitatively correct results here.

\section{XII}{Analytic Solution of the One-dimensional Equation}

We now discuss the analytical solution of the
{\Oned} {\FP} equation, (\oned).  We shall be mostly concerned with
the steady-state solution, since this is of the most
practical interest, and accurately approximates that given
by (\twod).  We will briefly consider the time scales given by
(\oned).  The methods we use and results we obtain
are very similar to those of Fisch,\ref{19, 20} who
treats the parallel {\Oned} {\FP}
equation.  Most of the analysis will be carried out for general
collision term, $C$.  We will however sometimes need to
know the functional dependence of $C$ on $v_\perp$.  We shall
therefore give the form of $C$ in the limit
$\vti \lsls v_\perp \lsls v_{te}$.  Using the forms for
$\nuab_\perp$ and $\nuab_\para$ given in Ref.\ 18 and assuming that
$T_e = T_i$, which is required in the derivation
of (\oned), and $\lambda_{ie} = \lambda_{ii}$
we have
$$C = \nzero \vti^2 {3\over 2} \left({\vti^3\over v_\perp^3}
+ {\vti^3\over v_0^3} \right), \eqn{\en{Ceq}{85}}$$
\xdef\Ceq{\eqprefix {85}}%
where $\nzero$ is given by (\nudef) and
$${v_0^3\over \vti^3} = {9Z_i\over 2(2/\pi)^{1/2}}
\left( {m_i\over m_e} \right)^{1/2}, \eqn{\en{vzerodef}{86}a}$$
or
$${v_0\over \vti} = \case{6.23,&\for \hjust{hydrogen},\cr
6.99,&\for \hjust{deuterium},\cr
7.48,&\for \hjust{tritium}.\cr}\eqn{\+b}$$
The contributions of $C_i$ and $C_e$ to $C$ are given
by the first and second terms in parentheses in (\Ceq).

The steady-state solution to (\oned) is\ref{20}
$$F(t \to \inf) = F_0 \exp-\int_{{\textstyle 0}}^{{\textstyle v_\perp}}
{v_\perp/\vti^2
\over 1+D/C}\,dv_\perp,\eqn{\en{Fss}{87}a}$$
\xdef\Fss{\eqprefix {87}}%
where $F_0$ is a normalization constant determined by
$$\int 2\pi v_\perp F\, dv_\perp = 1. \eqn{\+b}$$
If $D$ is only non-zero for $v_\perp^2 \grgr \vti^2$
then
$$F_0 \approx (2\pi\vti^2)^{-1}.\eqn{\en{Fzero}{88}}$$
\xdef\Fzero{\eqprefix {88}}%

The power dissipated in the steady state is given by (\pddef),
$$P = {2\pi n_0 m_i\over \vti^2} \int {Dv_\perp^3\over
1+D/C} F\,dv_\perp.\eqn{\en{Peq}{89}}$$
\xdef\Peq{\eqprefix {89}}%
This power dissipation is divided between the electron
and ion backgrounds according to
$$P_\beta = {2\pi n_0 m_i\over \vti^2} \int {C_\beta Dv_\perp^3\over
C+D} F\,dv_\perp,\eqn{\en{Pbeq}{90}}$$
\xdef\Pbeq{\eqprefix {90}}%
where $P_\beta$ is the power deposited in the background
species, $\beta$.

Equation (\Peq) may be evaluated in two limits.  Firstly,
in the collision dominated regime, for which $D \lsls C$
for all $v_\perp$, $F$ is approximately a Maxwellian and (\Peq)
becomes
$$P = {n_0 m_i\over \vti^4} \int Dv_\perp^3 \exp
\left(- {v_\perp^2\over 2\vti^2}\right)\,dv_\perp . \eqn{\en{pmax}{91}}$$
If $D(v_\perp) = 0$ for $v_\perp < v_1$ and $D(v_\perp)
\approx D_0$ (a constant) for $v_1 \le v_\perp
\lsapprox v_1 + \vti^2/v_1$, the integral may be approximated
by Laplace's method to give
$$P \approx n_0 m_i (v_1/\vti)^2 D_0 \exp(-\half v_1^2/\vti^2). \eqn{\e}$$
If $v_1^3 \lsls v_0^3$, nearly all of this goes to ion heating.

The second limit we shall consider is the ``plateau'' limit,
in which a plateau forms on $F$.  Suppose $D = 0$
for $v_\perp < v_1$ or $v_\perp > v_2$ (with $v_2 > v_1$)
and $D \grgr C$ otherwise.  If $D$ also satisfies
$$\int_{{\textstyle v_1}}^{{\textstyle v_2}}
{Cv_\perp\over D\vti^2}\,dv_\perp \lsls 1,
\eqn{\en{platcond}{93}}$$
then $F$ is approximately constant between $v_1$ and $v_2$.  Assuming
$v_1^2 \grgr \vti^2$, so that $F_0$ is given
by (\Fzero), (\Peq) becomes
$$P = {n_0 m_i\over \vti^4} \exp\left(-{v_1^2\over 2\vti^2}\right)
\int_{{\textstyle v_1}}^{{\textstyle v_2}}
Cv_\perp^3\,dv_\perp . \eqn{\e}$$
Substituting for $C$ from (\Ceq) we have
$$\eqalignno{P&=\fract{3}{2}n_0 m_i\nzero\vti[(v_2-v_1) +
\quarter(v_2^4-v_1^4)/v_0^3] \exp(-\half v_1^2/\vti^2), &(\e)\cr}$$
where the first and second terms in brackets are
the contributions of $P$ to the ion and electron heating
respectively.

The time scales may be readily computed for this case.  Since
the time scales derived from the {\Oned} equation are
not very accurate, we shall merely quote the results.  The
derivation closely follows that given in Ref.\ 19.  The
time for the distribution to flatten between $v_1$
and $v_2$ due to $D$ is
$$\tau_d = (v_2-v_1)^2/D. \eqn{\en{taud}{96}}$$
If $v_1(v_2-v_1) \grgr \vti^2$, the time scale for
particles to collisionally fill in the plateau is
$$\tau_c = \half(v_2^2-v_1^2)\vti^2/[v_1^2\,C(v_1)],\eqn{\e}$$
or substituting for $C$, (\Ceq), with
$v_1^3 \lsls v_0^3$
$$\tau_c = \fract 13(v_2^2-v_1^2)v_1/(\vti^3\nzero). \eqn{\en{tauc}{98}}$$
The time scale for the saturation of $P_c$ is $\tau_c$, while
$\tau_d$ is the time at which $P_d$ nearly reaches its steady-state
value.  For $t > \tau_d$ the power required for the
wave to accelerate the ions into the plateau decreases,
while the power to maintain them there increases.  Thus, $P_d$ is roughly
constant for $t > \tau_d$.

Many cases of interest will not be in the collisional or plateau
limits.  However, we may still obtain approximate results
by judiciously taking one or other limit.  For instance, if
we consider the case treated in the previous section, we see that
a plateau has formed over part of the range where $D$ is
finite, Fig.\ 14(b).  The plateau limit has some validity
in this case since $D$ cuts off quite sharply at the low
velocity end, Fig.\ 13, and so the height of the plateau is
fairly well determined.  We take $v_1$ to be given by
$$D(v_1) = C(v_1)\eqn{\e a}$$
and $v_2$ by
$$\int_{{\textstyle \omega/k_\perp}}^{{\textstyle v_2}}
{Cv_\perp\over D\vti^2}\,dv_\perp = 1.
\eqn{\+b}$$
(Thus, at $v_2$, $F$ is a factor, $e$, lower than the
plateau height.)  These give
$v_1 = 3.1\,\vti$ and $v_2 = 5.9\,\vti$, approximately.  We then
obtain
$$P \approx 5\times 10^{-2}\, n_0 m_i \vti^2 \nzero. \eqn{\e}$$
About 70\%\ of this goes to heating the ions.  Equation
(\+) is fairly close to the observed value for which
the numerial factor is $3.8 \times 10^{-2}$,
Fig.\ 15(a).  By far the most sensitive factor in determing
$P$ is $v_1$, since the height of the plateau depends
exponentially on $v_1$.  More accurate estimates of the
plateau height are possible.  However, if a more accurate
result that (\+) is required, it is probably easiest just to
evaluate (\Fss) and (\Peq) numerically.

With these values of $v_1$ and $v_2$ the time scale become
$\tau_d = 5/\nzero$ [taking $D$ to be given by (\dval)] and
$\tau_c = 26/\nzero$.  These agree quite well with the
results of the numerical solution of (\oned) given in
Fig.\ 15(b).  However, as we noted these times (and, in
particular, $\tau_c$) differ from the times given by the {\Twod} {\FP}
equation.

\section{XIII}{Discussion}

We have computed the rate of diffusion of ions in perpendicular
velocity space in the presence of a uniform magnetic
field and a perpendicularly
propagating electrostatic wave for which $\omega \grgr \Omega_i$.  The
diffusion coefficient is given by (\Dun).  From the
results of I\null, it is seen that
this diffusion coefficient also applies to diffusion in a
lower hybrid wave (for which $k_\para$ is finite, but
$k_\para \lsls k_\perp$) in a weakly inhomogeneous
magnetic field.  The effect of nonuniformity of the lower hybrid wave
is modeled by introducing the geometrical factor, $\gamma$, into
(\Dun).

In order to determine the heating rate, collisions are included,
and a {\Oned} {\FP} equation, (\oned), is derived.  From this the
steady-state power dissipation, $P$, may be
found, (\Peq).  Provided $\omega/(k_\perp \vti)$ becomes
suffi\-cient\-ly small (on the order of 4) this
may result in strong damping of the lower hybrid wave
and efficient heating of the ions.  (See the example
in Sec.\ XI\null.)

These results may be used in a simulation of lower hybrid heating
with a transport code, provided the time scale for the
evolution of the plasma exceeds $\tau_c$.  The quantities,
$P_e$ and $P_i$, (\Pbeq), which represent the powers deposited
per unit volume, would then appear as source terms in the
equations for the electrons and ions.  The wave would
obey an evolution equation
$$v_{gr}\, \pd W(r, t)/\pd r = - P(r, t)/\gamma, \eqn{\e}$$
where $W$ is the energy density of the wave, $r$ is the
minor radius of the tokamak, and $v_{gr}$ is the radial group
velocity.  In this equation $t$ is a parameter.  $P$ and $W$
would vary with time as the background distributions evolve.

Lastly we consider the question of particle transport due to
stochastic heating.  Firstly, there is the change in the transport
coefficients due to the altered distribution function of the ions,
Fig.\ 14.  Since the heating initially takes place in the
perpendicular direction, this would lead to an increase in
the number of trapped ions, which may have a deleterious effect
on the transport.  Secondly, the equations of motion, (\lfl),
give a shift in the guiding centers of the ions.  The
amount of this shift is most easily calculated from the
Hamiltonian, (A1), since $-I_2$ is the $x$ guiding center
position.  (There is no shift in the $y$ guiding center.)  Thus,
we have $-\Delta I_2 = \Delta I_1/\nu$ (after averaging over a
cyclotron period) or, undoing the normalizations
$$\Delta\vec{x} = (k_\perp/\omega)\,\Delta\mu\,\^{\vec{k}}\cross
\^{\vec{B}},\eqn{\e}$$
where $\mu$ is the magnetic moment,
$\half m_i v_\perp^2/(q_i B_0)$.  When considering heating by the
injected lower hybrid wave (rather than parametric decay products)
$\vec{k}$ is mainly in the radial direction and so $\Delta \vec{x}$
is mostly in the poloidal direction.  So this effect does
not lead to any outward transport of ions.

\acknowledgments

The author wishes to thank N. J. Fisch, J. M. Greene,
J. A. Krommes, and A. B. Rechester for useful discussions.

This work was supported by the U. S. Department of Energy
under Contract No.\ EY--76--C--02--3073.

\def\secflag{A}\appendix{A}{Difference Equations from the Hamiltonian}

In this appendix we derive the difference equations, (\de),
from the Hamil\-tonian for the ion.  Using this approach
it is easy to establish what approximations are made.  In I it
was shown that the motion of an ion in the fields,
(\fields), is governed by the Hamiltonian,
$$\eqalignno{H&=I_1 + \nu I_2 -\a\sin[\Ihalf\sin w_1 - w_2]
\cr\eqskp
&=I_1 + \nu I_2 -\a\sum_{m=-\inf}^\inf J_m[\Ihalf]\sin(mw_1-w_2),
&(\en{Heq}{1})\cr}$$
\xdef\Heq{\eqprefix {1}}%
where the normalizations are the same as used in
(\lfl), $J_m$ is the Bessel function of order $m$,
$y = \Ihalf\sin w_1$, and $x = -I_2 - \Ihalf\cos w_1$.  The
action variable, $I_1$, is related to the Larmor radius,
$r$ (\pdef c), by $I_1 = \half r^2$.  We now perform a series
of transformations, similar to those made in Appendix
C of I\null, to bring the sum in (\+) into the form of an
infinite sum of the product of two trigonometric terms.  Hamilton's
equations of motion may then be written in terms of delta
functions.

We first transform (\+) to hatted variables using the
generating function,
$$F_2 = \^I_1(nw_1 - w_2) + \^I_2 w_2, \eqn{\e}$$
where $n$ is given by (\ddef).  Equation (\Heq) then
becomes
$$\eqalignno{\^H &= -\d \^I_1 + \nu \^I_2 -\a\sum_m J_m[\Inhalf]
\sin\left[{m\over n}\^w_1 - \left(1-{m\over n}\right)\^w_2\right].
&(\e)\cr}$$
Since $\^w_1$ is slowly varying, whereas $\^w_2$ is still
rapidly varying (assuming $n \grgr 1$) the dominant contributions
to the sum in (\+) come from $m \approx n$.  Recognizing this we
replace $m$ by $n + k$ in the sum to obtain
$$\eqalignno{\^H &= -\d \^I_1 + \nu \^I_2 -\a\sum_{k = -N}^N J_{n+k}[\Inhalf]
\sin\left[\left(1+{k\over n}\right)\^w_1 + {k\over n}\^w_2\right],
&(\en{Hhat}{4})\cr}$$
\xdef\Hhat{\eqprefix {4}}%
where $N$ is some integer satisfying $1 \lsls N \lsls n$.  [It
is convenient to choose $N\approx\nuth$.]  As a result of this
inequality we approximate the factor $(1 + k/n)$ by unity.  We
next use the large argument expansion of the Bessel function
$$\eqalignno{J_m(r) &= (2/\pi)^{1/2} (r^2-m^2)^{-1/4}\cr\eqskp
&\qquad\times \cos[(r^2-m^2)^{1/2} - m\cos^{-1}(m/r) -\pi/4] &(\e)\cr}$$
[Ref.\ 3, Eq.\ (9.3.3)], which is valid for $r > m +
(\half m)^{1/3}$.  In order to put the Hamiltonian
into the required form we need to be able to do three
things:  approximate all the Bessel functions in
(\Hhat) using (\+), which requires $r > (n+N) + (\half n)^{1/3}$;
replace the amplitude factor $(r^2-m^2)^{-1/4}$ by
$(r^2-\nu^2)^{-1/4}$, which requires $r^2-n^2 \grgr Nn$;
and to Taylor-expand the argument of the cosine in (\+)
in $k$ about $k = \d$, keeping only the first two terms,
which requires $r^2-n^2 \grgr N^4$.  With $N \approx \nuth$
all these inequalities become
$r-\nu \grgr \nuth$.  The Bessel functions
in (\Hhat) then become
$$J_{n+k}(r) = (2/\pi)^{1/2} (r^2-\nu^2)^{-1/4}
\cos(R-k\phi+\d\phi),\eqn{\e}$$
where $r = \Inhalf$, $R = g(r) \eqv (r^2-\nu^2)^{1/2}
- \cos^{-1}(\nu/r) - \pi/4$, and $\phi = \cos^{-1}(\nu/r)$.  We shall
regard the $r$'s appearing in the amplitude factor
and in the definition of $\phi$ as being parameters.  This
requires that neither of these quantities change significantly
when $r$ changes by a period of the Bessel function,
$2\pi(\pd R/\pd r)^{-1}$.  The conditions on $r$ and $\nu$
already assumed are sufficient to guarantee this.  In
addition, if $\a$ exceeds the stochasticity threshold,
$r$ may change by more than the period of the Bessel
function, in which case we must re-examine our assumptions
about the constancy of the amplitude factor and $\phi$.  We
shall return to this point after we have derived the
difference equations.

We now transform to a new set of variables, distinguished by
tildes, using the generating function,
$$F_3 = - G(\^I_1)\~w_1 - \^I_2\~w_2, \eqn{\e}$$
where $G(\^I_1) \eqv g[\Inhalf]$.  This gives
$$\~I_1 = G(\^I_1),\qquad \^w_1 = G^\prime(\^I_1)\~w_1.\eqn{\e}$$
We solve both of these locally by expanding $\^I_1$ about
some point $\^I_{10}$ and keeping only the terms involving
$G^\prime(\^I_{10})$.  Thus, we obtain
$$\^I_1 = \~I_1/G^\prime(\^I_{10}) + \hjust{const.},\qquad
\^w_1 = G^\prime(\^I_{10})\~w_1. \eqn{\e}$$
This is consistent with the approximations already made.  Writing
$G^\prime(\^I_{10})$ in terms of the Larmor radius, $r$
(which, like $\^I_{10}$, we regard as constant), we have
$$G^\prime(\^I_{10}) = g^\prime(r)\,\pd r/\pd \^I_1
\approx (r^2-\nu^2)^{1/2}\nu/r^2 \eqv Q. \eqn{\e}$$
The Hamiltonian becomes
$$\eqalignno{\~H &= -{\d \over Q} \~I_1
+ \nu\~I_2\cr\eqskp
&\qquad\null-{\a(2/\pi)^{1/2}\over (r^2-\nu^2)^{1/4}}
\sum_k \cos(\~I_1 - k\phi + \d\phi)
\sin\left(Q\~w_1 + {k\over n}\~w_2\right).
&(\e)\cr}$$

Finally, we perform the scaling transformation,
$$\eqaligntwo{M&=Q\~H,&&&(\e a)\cr\eqskp
J_1&=\~I_1,& \psi_1 &= Q\~w_1, &(\+b)\cr\eqskp
J_2&= Q\nu \~I_2,& \psi_2 &= \~w_2/\nu, &(\+c)\cr}$$
so that the Hamiltonian
is given by
$$\eqalignno{M &= -\d J_1 + J_2 - A\sum_{k=-\inf}^\inf \cos(J_1 - k\phi + \d\phi)
\sin(\psi_1 + k\psi_2), &(\e)\cr}$$
where $A$ is given by (\Adefp) and we have approximated $\nu/n$
by unity to give the last term in the argument to the sine.  We
have extended the limits of the sum to infinity;  the additional
terms introduced are non-resonant, and so do not have much effect.

Hamiltons's equations now give $\.\psi_2 = 1$ or $\psi_2 = t$
and
$$\eqalignno{\.\psi_1&=-\d + A \suml_k \sin(J_1 - k\phi + \d\phi)
\sin(\psi_1 + kt), &(\e a)\cr\eqskp
\.J_1&= A \suml_k \cos(J_1 - k\phi + \d\phi)
\cos(\psi_1 + kt). &(\+b)\cr}$$
Defining new variables,
$$\eqaligntwo{\gamma &= n\pi-\psi_1,& \r &= J_1+\nu\pi,&(\en{foo}{15}a)\cr\eqskp
u &= \gamma-\r,& v &= \gamma+\r,&(\+b)\cr}$$
\xdef\foo{\eqprefix {15}}%
we find
$$\eqalignno{
\.u &= \d-A\suml_k\cos[v-(\pi-\phi)\d-k(t+\phi)]\cr\eqskp
&= \d-2\pi A\cos[v-(\pi-\phi)\d]\sum_{j=-\inf}^\inf\^\d(t+\phi-2\pi j),
&(\e a)\cr\eqskp
\.v &= \d+A\suml_k\cos[-u-(\pi-\phi)\d+k(t-\phi)]\cr\eqskp
&= \d+2\pi A\cos[u+(\pi-\phi)\d]\sum_{j=-\inf}^\inf\^\d(t-\phi-2\pi j).
&(\+b)\cr}$$
We have used the notation, $\^\d$, to denote the
Dirac delta function (to distinguish it from the variable,
$\d$) and we have used the identity,
$$\sum_{k=-\inf}^\inf \cos kt = 2\pi\sum_{j=-\inf}^\inf
\^\d(t - 2\pi j). \eqn{\e}$$
Defining $u_j = u(t = 2\pi j-\pi)$ and similarly for $v$,
we may obtain
$$\eqalignno{
u_{j+1}-u_j &= 2\pi\d-2\pi A\cos v_j,&(\e a)\cr\eqskp
v_{j+1}-v_j &= 2\pi\d+2\pi A\cos u_{j+1}.&(\+b)\cr}$$
It is readily established that $\r$ as given by (\foo a)
agrees with (\vdef a), and that $\gamma$ at $\psi_2 = 2\pi j
- \pi$ is equal to $w_2$ at $w_1 = 2\pi j -\pi$ and so is
equal to $\t_j$ as defined in (\vdef b).  Thus, the difference
equations given in (\+) agree with those derived in Sec.\ II\null,
(\de).  The final condition on the validity of
(\+) is that the change in $\r$ in one iteration is
insufficient to cause an appreciable change in $A$ through
$r$.  (This then justifies taking $A$ to be a constant.)  The
maximum change in $\r$ in one iteration is on the order
of $A$.  Therefore the fractional change in $A$ due to this
change in $\r$ is
$$A{\pd\log A\over \pd \r} = {\pd A\over \pd \r}
= {\pd A\over \pd r}{\pd r\over \pd \r} \sim
{\a \nu\over (r^2-\nu^2)^{5/4}},\eqn{\e}$$
where, in differentiating $A$ we have only considered its
dependence on $r$ through $(r^2-\nu^2)^{1/4}$.  Demanding
that (\+) be small gives
$$\a \lsls (r^2-\nu^2)^{5/2}/\nu\quad\hjust{or}\quad
A \lsls (r^2-\nu^2)^{3/2}/r^2.\eqn{\e}$$
The other assumptions made are $\nu \grgr 1$ and $r-\nu
\grgr \nuth$.

\appendix{B}{First-Order Accelerator Modes}

We consider here the simplest of the accelerator modes.  Such
modes are island systems around first-order fixed points in the
$(u, v)$ plane;  i.e., points for which $u_1-u_0 = -2\pi n$
and $v_1-v_0 = 2\pi m$, where $m$ and $n$ are integers.  The
amount by which $\r$ increases per iteration is $s\pi$,
where $s = m+n$.  From (\de) the fixed points are given by
$$\d+A\cos u_0 = m,\qquad \d-A\cos v_0 = -n.\eqn{\en{fix}{1}}$$
\xdef\fix{\eqprefix {1}}%
In order for an island to exist, the fixed point, $(u_0, v_0)$,
must be elliptic.  The stability of $(u_0, v_0)$ is determined
by the eigenvalues of the linearized mapping.  From (\Jdef)
we have
$$\mat{J}_0 = \twocol{1&2V\cr -2U&1-4UV\cr},\eqn{\e}$$
where $U = \pi A\sin u_0$ and $V = \pi A \sin v_0$.  The
eigenvalues are given by
$$\lambda = 1-2UV \pm 2(U^2V^2 - UV)^{1/2} . \eqn{\e}$$
The fixed point is elliptic if $\lambda$ is complex,
which requires $0 < UV < 1$.  Sub\-sti\-tut\-ing for $\sin u_0$
and $\sin v_0$ gives
$$\eqalignno{A^2-(m-\d)^2 &> 0, &(\en{range}{4}a)\cr\eqskp
A^2-(n+\d)^2 &> 0, &(\+b)\cr\eqskp
[A^2-(m-\d)^2][A^2-(n+\d)^2] &< 1/\pi^4, &(\+c)\cr}$$
\xdef\range{\eqprefix {4}}%
as the conditions for first-order accelerator modes.  When
$\abs{A}$ just exceeds $\max\penalty100(\abs{m-\penalty100\d},
\penalty50\abs{n+\d})$ then
(\+) is satisfied.  At this point elliptic and
hyperbolic fixed points appear (with different signs
for $\cos u_0$ or $\cos v_0$).  As $\abs{A}$ is increased
further (\+c) fails and the elliptic fixed point turns
into a hyperbolic fixed point with reflection.  Thus,
(\+) defines a range, $A_\l < \abs{A} < A_u$, which
is a function of $\d$, $m$, and $n$, in which
first-order elliptic fixed points exist.

Figure 16 shows the behavior of the mode with $m = 1$
and $n = 0$ for $\d = \half$.  From (\+)
we see that for these parameters $A_\l = \half$ and
$A_u = (\quarter + \pi^{-2})^{1/2} = 0.5927$.  At
$A = 0.55$, Fig.\ 16(a), the first-order island is visible.  This
is surrounded by higher-order (4th and 32nd) accelerator modes.  When
$A$ just exceeds $A_u$, Fig.\ 16(b), the center of the
first-order island becomes unstable, becoming a hyperbolic
fixed point, separating two second-order islands.  However,
although this point is unstable, it is confined to the
island system, since there are still eigencurves with
the original topology surrounding the system of two
islands and one hyperbolic fixed point.  Thus, a
point started close to the hyperbolic point will be forever
accelerated.  These eigencurves are soon destroyed, however,
allowing the escape of particles starting near the fixed point.  In
Fig.\ 16(c) we see that the particle spends on the order of $1000$
iterations encircling the second-order islands before
escaping.  As $A$ is increased further, the second-order
islands rapidly decrease in size, becoming unstable when
$A$ is slightly greater that 0.601.

From Fig.\ 16 we see that we should study accelerator modes of many
different orders in order to have a complete understanding
of their effect on the diffusion of particles.  However,
while it is possible to thoroughly study first-order
fixed points, just cataloguing the higher-order fixed points
becomes a monumental task for $A \grapprox 1$.  For example,
there are over 100 second-order fixed points present
for $A = 1$.  We will therefore pursue only the first-order
fixed points.

In Fig.\ 17 we show the range in $\abs{A}$ defined
by (\+) as a function of $\d$ with $s = 0$, 1, and
2.  (For each value of $s$, all values of $m$
and $n$ satisfying $m+n=s$ are considered.)  The pattern evident
in these figures is repeated for higher values of
$s$.  From (\+) and Fig.\ 17 we see that an accelerator mode,
with a given value of $s$, first appears at
$$A_\l = \case{\abs{s/2} + \abs{\d},&\for s \hjust{ even},\cr\caseskp
\abs{s/2} + \half - \abs{\d},&\for s \hjust{ odd}.\cr}\eqn{\en{bar}{5}}$$
\xdef\bar{\eqprefix {5}}%
(Recall that $\d$ is defined such that $\abs{\d}
\le \half$.)

The accelerator mode disappears at a slightly higher amplitude,
but periodically reappears (with different $m$ and $n$,
but the same $s$) at
$$A_\l = \case{\abs{s/2} \pm \abs{\d} + p,&\for s \hjust{ even},\cr\caseskp
\abs{s/2} \pm (\half - \abs{\d}) + p,&\for s \hjust{ odd},\cr}
\eqn{\en{foo}{6}}$$
\xdef\foo{\eqprefix {6}}%
where $p$ is a positive integer.  For a given $s$, the gap in
which the accelerator mode exists, $\Delta A \eqv
A_u - A_\l$, is widest when $A_\l = \abs{s/2}$, $\d = 0$ for $s$
even and $\abs{\d} = \half$ for $s$ odd.  In this case we have
$$A_u = \left({s^2\over 4}+{1\over \pi^2}\right)^{1/2}.
\eqn{\e}$$
For $s=0$ (i.e., no acceleration), we have $A_\l = 0$
and $A_u = 1/\pi$.  This is the range in which the central
fixed points of the islands in Fig.\ 3(a) are stable.  If
$\abs{s} \grgr 1$ then
$$\Delta A = {1\over 2\pi^2 A_\l}. \eqn{\en{deltab}{8}}$$
\xdef\deltab{\eqprefix {8}}%
When $\abs{A}$ greatly exceeds the minimum threshold
for a given $s$, so the $A_\l$ is given by (\foo)
with $p \grgr 1$, we may find an approximate solution for
$\Delta A$ from (\range):
$$\Delta A = \case{{1\over 2\pi^2A_\l},&\for s=0,\cr\caseskp
{1\over 4\pi^4sA_\l^2},&\for s\ne 0.\cr}\eqn{\en{deltaa}{9}}$$
\xdef\deltaa{\eqprefix {9}}%

Since the number of first-order accelerator modes increases with
increasing $A$ (see Fig.\ 17), we are lead to ask how the
fraction of phase space occupied by accelerator modes varies
with increasing $A$.  We try to find an upper bound
on this quantity for the first-order modes.  We estimate the
size of an accelerator mode by noting that the
maximum radius of the island is given by the distance between
the elliptic and hyperbolic fixed points.  We
consider here the pair of fixed points that are born
together and we assume that the islands are
roughly circular.  From (\fix) this distance is about
$[(A-A_\l)/A_\l]^{1/2}$ (ignoring numerical factors).  This
is maximaized by letting $A = A_u$.  The maximum
area of the island is then $\Delta A/A_\l$.  For a given
$A$, modes with $s$ from 1 to about $2A$ can be present;  thus,
the total area of the accelerator modes is on the order of
$$\sum_{s=1}^{2A}{\Delta A\over A_\l} \sim
\sum_{s=1}^{2A}{1\over sA^3} \sim {\log A\over A^3}.\eqn{\e}$$
Here we have used the form of $\Delta A$ for $p \grgr 1$ and
$s \ne 0$, (\deltaa).  If $\d = 0$ or $\half$ then there is an
additional contribution of order $A^{-2}$ from the mode that just
appeared for the first time by satisfying (\bar); in this
case $\Delta A$ is given by (\deltab).  This then dominates
over (\+).  In either case the area of the accelerator
modes decreases with $A$.

However, the modes tend to line up close to
$\cos u = 0$ and $\cos v = 0$, since (\range a), say, may
be well satisfied, when (\range b) becomes true.  We should
investigate whether the modes can form a barrier inhibiting
the diffusion of particles.  The sum of the widths of the
accelerator modes is
$$\sum_{s=1}^{2A} (sA^3)^{-1/2} \sim A^{-1}, \eqn{\e}$$
which also decreases with $A$.

We conclude that the effect of the first-order
accelerator modes decreases with increasing $A$.  This
leaves open the possibility that the effect of all the
accelerator modes may not decrease.  We have, however, observed
no evidence of this.

\appendix{C}{Monte Carlo Solution}

We describe here the solution of (\diffnorm) by a Monte Carlo
method.  We first define a new velocity coordinate,
$\mu = \half r^2$, so that (\diffnorm) becomes
$${\pd f\over \pd t}={\pd\over \pd\mu}r^2D{\pd\over \pd\mu}f.
\eqn{\en{mueq}{1}}$$
\xdef\mueq{\eqprefix {1}}%
(The variable $\mu$ is proportional to the magnetic
moment and is the same as the action, $I_1$,
used in Appendix A.)  We solve (\mueq) by dividing each
cyclotron period up into $M$ equal time intervals of $2\pi/M$.  The
distribution function is recorded  by the positions
of $N$ particles, $\mu_{i,j}$, where the subscript, $i$, refers to
the particle number and $j$ to the time step, $j = Mt/2\pi$.  The
distribution function, $f$, is found by converting $\mu$ back to $r$ and
letting
$$f(r) = {\d N\over 2\pi Nr\,\d r}, \eqn{\e}$$
where $\d N$ is the number of particles with velocity
between $r - \d r/2$ and $r + \d r/2$.  We advance $f$ by the
difference equation,
$$\mu_{i,j+1}-\mu_{i,j}=\case{
+\dP,&\hjust{with probability }\half,\cr\caseskp
-\dM,&\hjust{with probability }\half.\cr}\eqn{\en{muadv}{3}}$$
\xdef\muadv{\eqprefix {3}}%
[Do not confuse $\d^\pm$ with the parameter, $\d$, defined in
(\ddef).]  In the limit,
$M \to \inf$, both $\dP$ and $\dM$ are given by $\d(\mu)$,
where
$$\d(\mu) \eqv (2\pi r^2D/M)^{1/2}.\eqn{\en{deldef}{4}}$$
\xdef\deldef{\eqprefix {4}}%
However, there are corrections to the expressions of $\dP$
and $\dM$ of order $M^{-1}$ which must  be included if we
are to correctly model (\diffnorm).  [If we do not include
these corrections we end up solving the equation,
$\pd f/\pd t = (\pd^2/\pd\mu^2) r^2Df$.]  We derive these
corrections by demanding that the steady-state solution given
by the (\muadv) agrees with
the steady-state solution to (\mueq), viz., $f =
\hjust{const}$.  Letting $f$ be a constant (which we take to be
unity), the flux of particles, $S$, at $\mu$ is
$$S(\mu) =\half[\dP(\mu-\epsilon^+)-\dM(\mu+\epsilon^-)],
\eqn{\en{flux}{5}}$$
\xdef\flux{\eqprefix {5}}%
where
$$\epsilon^+ =\dP(\mu-\epsilon^+),\qquad
\epsilon^- = \dM(\mu+\epsilon^-).\eqn{\e}$$
Equation (\flux) just expresses the fact that half the particles
between $\mu-\epsilon^+$ and $\mu$ will pass
the point $\mu$ is one time step, and similarly for the particles
between $\mu$ and $\mu+\epsilon^-$.  Setting $S(\mu)$ to
zero gives
$$\dM(\mu+\epsilon)=\dP(\mu-\epsilon)=\epsilon. \eqn{\en{foo}{7}}$$
\xdef\foo{\eqprefix {7}}%
Putting $\epsilon = \d(\mu)$ we recover the correct result for
$\dP$ and $\dM$ in the limit, $M \to \inf$.  This gives
$$\dM[\mu+\d(\mu)]=\dP[\mu-\d(\mu)]=\d(\mu). \eqn{\en{stepeqp}{8}}$$
\xdef\stepeqp{\eqprefix {8}}%
Equation (\+) has a very simple graphical representation which is
shown in Fig.\ 18.  In the limits, $M \to \inf$ and $\d \to
0$, we can make a Taylor series expansion about $\mu$ to give
$$\dP(\mu)-\dM(\mu) = d(\d^2)/d\mu. \eqn{\e}$$
This acts as a friction force which appears in the diffusion
equation, (\mueq), when we write it in the form,
$${\pd f\over \pd t}=-{\pd\over \pd\mu}\left[
\left({\pd r^2D\over \pd\mu}\right)f
-{\pd\over \pd\mu}(r^2Df)\right].\eqn{\e}$$

From Fig.\ 18 we see that $\dP$ or $\dM$ becomes double-valued
if $\abs{d\d/d\mu} > 1$.  This can always be prevented
by choosing $M$ large enough.  However, from
the definition of $D$, (\dnorm), we see that $D$ has quite large
gradients for $r < \nu-\sqrt{\a}$.  We avoid choosing $M$
to be large to compensate for this by defining
$\dP(\mu)$ to be $\d(\mu)$ and basing the definition
of $\dM$ on $\dP$ with
$$\dM[\mu + 2\dP(\mu)] = \dP(\mu) = \d(\mu), \eqn{\en{stepeq}{11}}$$
\xdef\stepeq{\eqprefix {11}}%
which is obtained by letting $\epsilon = \d(\mu-\epsilon)$
in (\foo).  In the limit, $M \to \inf$,
(\+) agrees with (\stepeqp).

To summarize:  We solve (\mueq) and hence (\diffnorm) using
the difference scheme, (\muadv), with $\dP$ and $\dM$ defined
by (\stepeq) and (\deldef).  $M$ is chosen to be 25.

\references
\refno{1}C. F. F. Karney, Phys.\ Fluids \vol{21}, 1584 (1978).
\refno{2}Y. Gell and R. Nakach, in {\ital Plasma Physics and
Controlled Nuclear Fusion Research 1978},
(International Atomic Energy Agency, Vienna),
paper IAEA--CN--37--G--1 (to be published).
\refno{3}M. Abramowitz and I. A. Stegun (editors),
{\ital Handbook of Mathematical Functions},
(U. S. Government Printing Office, Washington, D.C., 1964).
\refno{4}C. F. F. Karney, Ph.D. thesis,
Massachusetts Institute of Technology (1977).
\refno{5}A. V. Timofeev, Nucl.\ Fusion \vol{14}, 165 (1974).
\refno{6}The Mathlab Group, {\ital MACSYMA Reference Manual}, Version 9,
Laboratory for Computer Sci\-ence, Massachusetts Institute of
Technology (1977).
\refno{7}M. V. Berry, in {\ital Topics in Nonlinear Dynamics},
edited by S. Jorna, Am.\ Inst.\ Phys.\ Conf.\ Proc., No.\ 46
(American Institute of Physics, 1978), p.\ 16.
\refno{8}J. Ford, in {\ital Fundamental Problems in Statistical
Mechanics}, edited by E. D. G. Cohen (North-Holland, Amsterdam, 1975),
Vol.\ 3, p.\ 215.
\refno{9}B. V. Chirikov, Phys.\ Repts.\ \vol{52}, 263 (1979).
\refno{10}V. I. Arnold, {\ital Mathematical Methods of Classical
Mechanics}, translated by K. Vogtmann and A. Weinstein (Springer-Verlag, 1978).
\refno{11}J. M. Greene, J.\ Math.\ Phys.\ \vol{20}, 1183, (1979).
\refno{12}A. Fukuyama, H. Momota, R. Itatani, and T. Takizuka,
Phys.\ Rev.\ Lett.\ \vol{38}, 701 (1977).
\refno{13}S. Chandrasekhar, Rev.\ Mod.\ Phys.\ \vol{15}, 1 (1943).
\refno{14}M. C. Wang and G. E. Uhlenbeck,
Rev.\ Mod.\ Phys.\ \vol{17}, 323 (1945).
\refno{15}E. Lazarro, Lettere al Nuovo Cimento \vol{22}, 625 (1978).
\refno{16}R. J. Briggs and R. R. Parker, Phys.\ Rev.\ Lett.\ \vol{29},
852 (1972).
\refno{17}B. A. Trubnikov, in {\ital Reviews of Plasma Physics},
edited by M. A. Leontovich (Con\-sul\-tants Bureau, N.Y., 1965),
Vol.\ 1, p.\ 105.
\refno{18}D. L. Book, {\ital NRL Plasma Formulary},
Naval Research Lab.\ (1978).
\refno{19}N. J. Fisch, Ph.D.\ thesis, Massachusetts Institute of
Technology (1978).
\refno{20}N. J. Fisch, Phys.\ Rev.\ Lett.\ \vol{41}, 873 (1978).
\refno{21}C. F. F. Karney and N. J. Fisch, Phys.\ Fluids \vol{22},
1810, (1979).

\vfill\penalty-10000
\Fig. 1.  (792078:0.6)
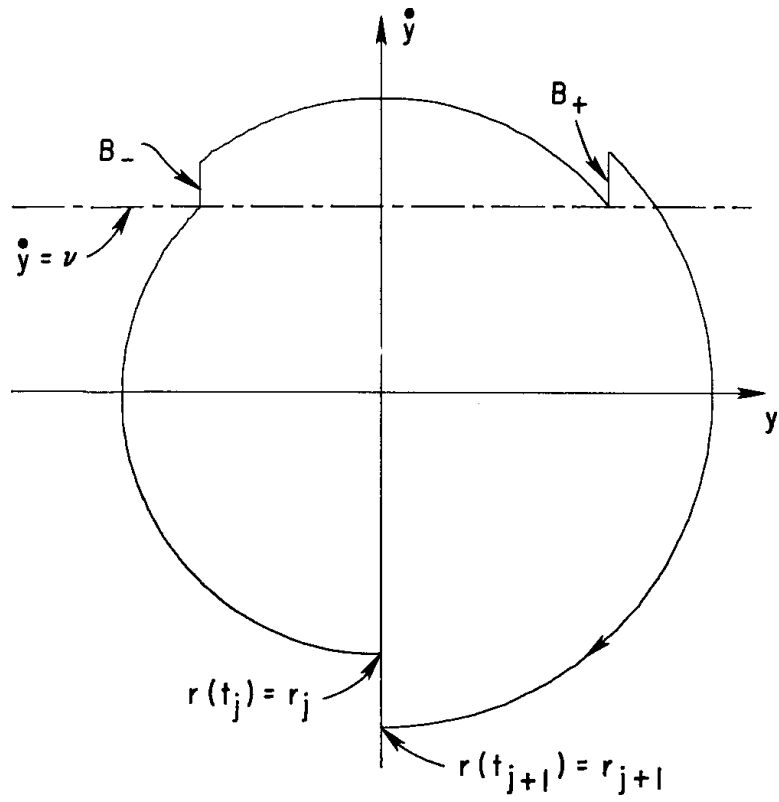
Motion of an ion in velocity space, showing the kicks
it receives when passing through wave-particle resonance.
\Fig. 2.  (792087:0.75)
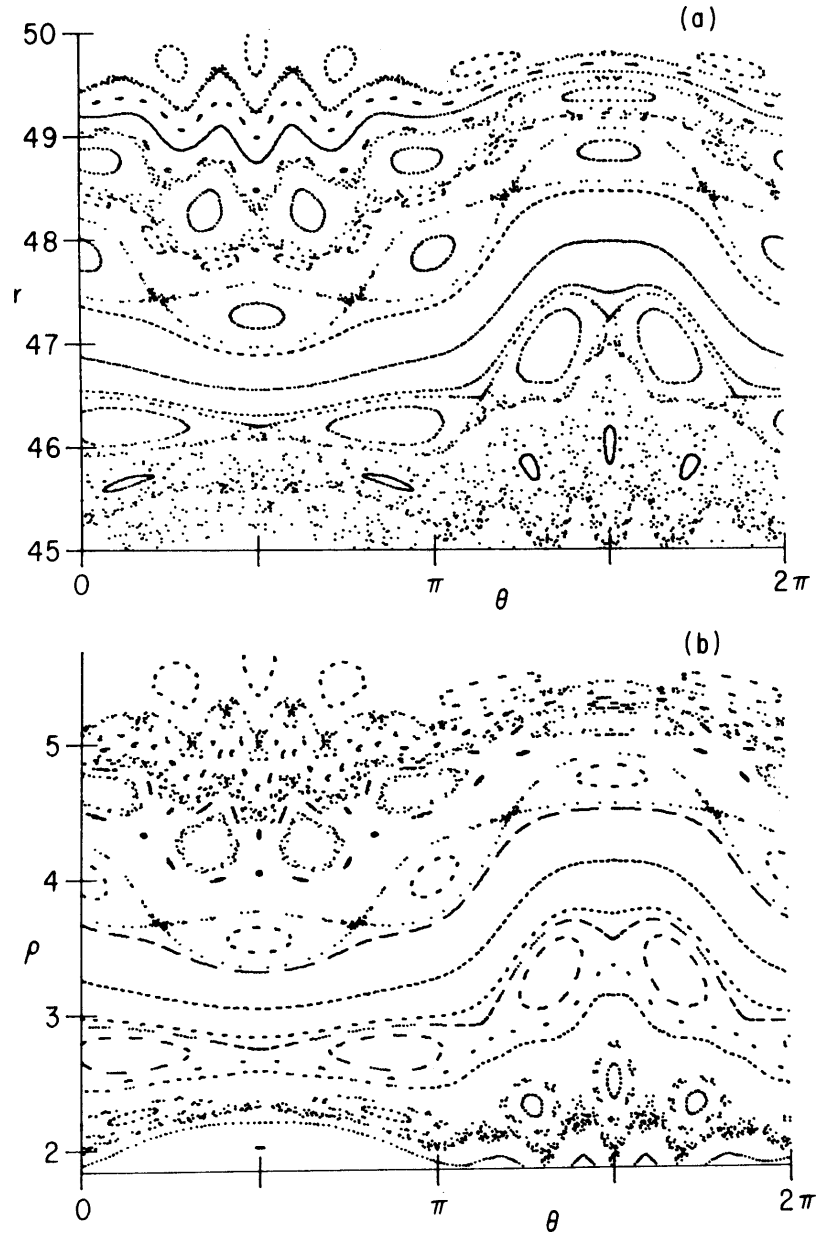
Comparison of the difference equations, (\de),
with the Lorentz force law, (\lfl).  (a) The mapping of the
$(\t, r)$ plane using (\lfl) with $\nu = 30.23$
and $\a = 2.2$.  [This is taken from
Fig.\ 3(b) of \hjust{I}.]  (b) The mapping of the $(\t, \r)$ plane
under $T$ using (\de), with $\d = 0.23$ and $A = 0.1424$, which
is given by (\Adefp) with $\nu = 30.23$, $\a = 2.2$, and
$r = 47.5$.  In each case the trajectories of 24 particles are followed for
300 orbits.  Thus, in (b) the points, $T^j(\t_0, \r_0)$ for
$0 \le j \le 300$, are plotted for 24 different initial conditions,
$(\t_0, \r_0)$.
\Fig. 3.  (792080:0.75)
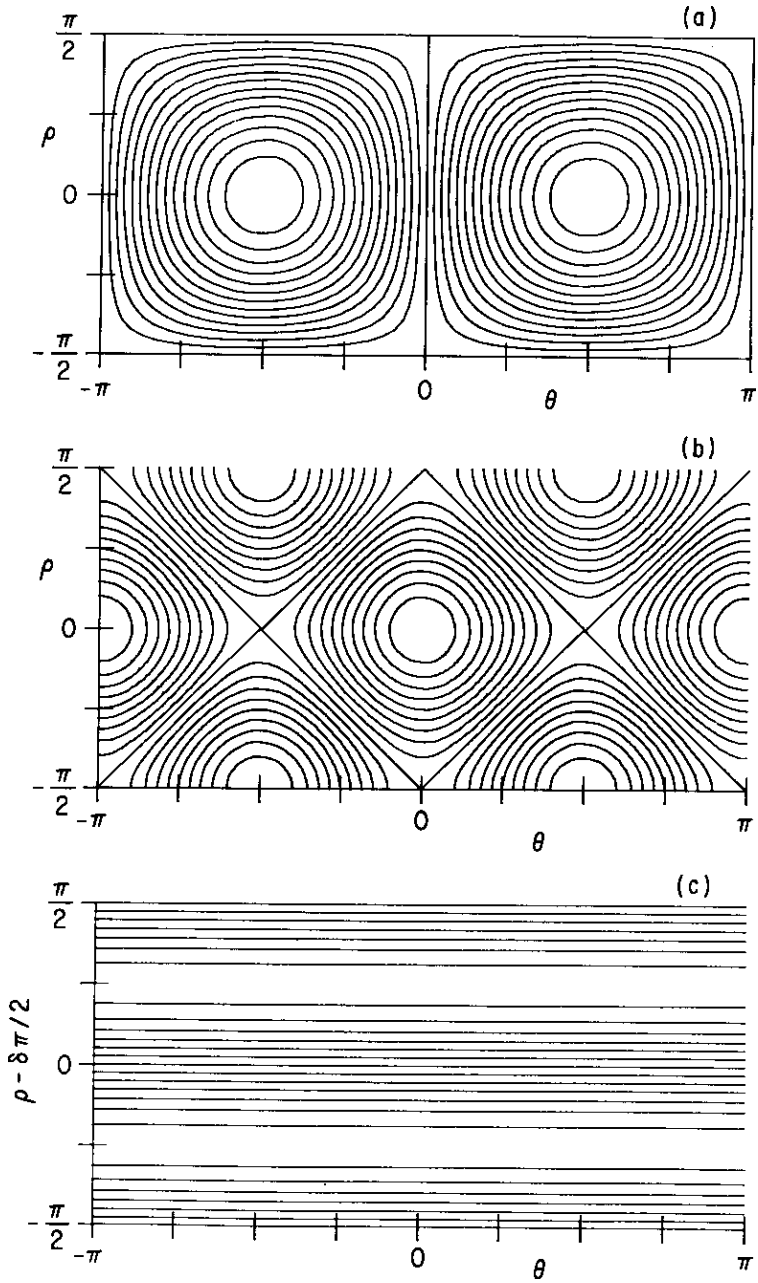
Trajectories in the $(\t, \r)$ plane for small $A$ and
(a) $\d = 0$, (b) $\d = \half$, (c) $\d = s/p$ (not equal to 0
or $\half$).  The trajectories are obtained by plotting curves for
which the Hamiltonian, $h$, is constant.
\Fig. 4.  (792086:0.7)
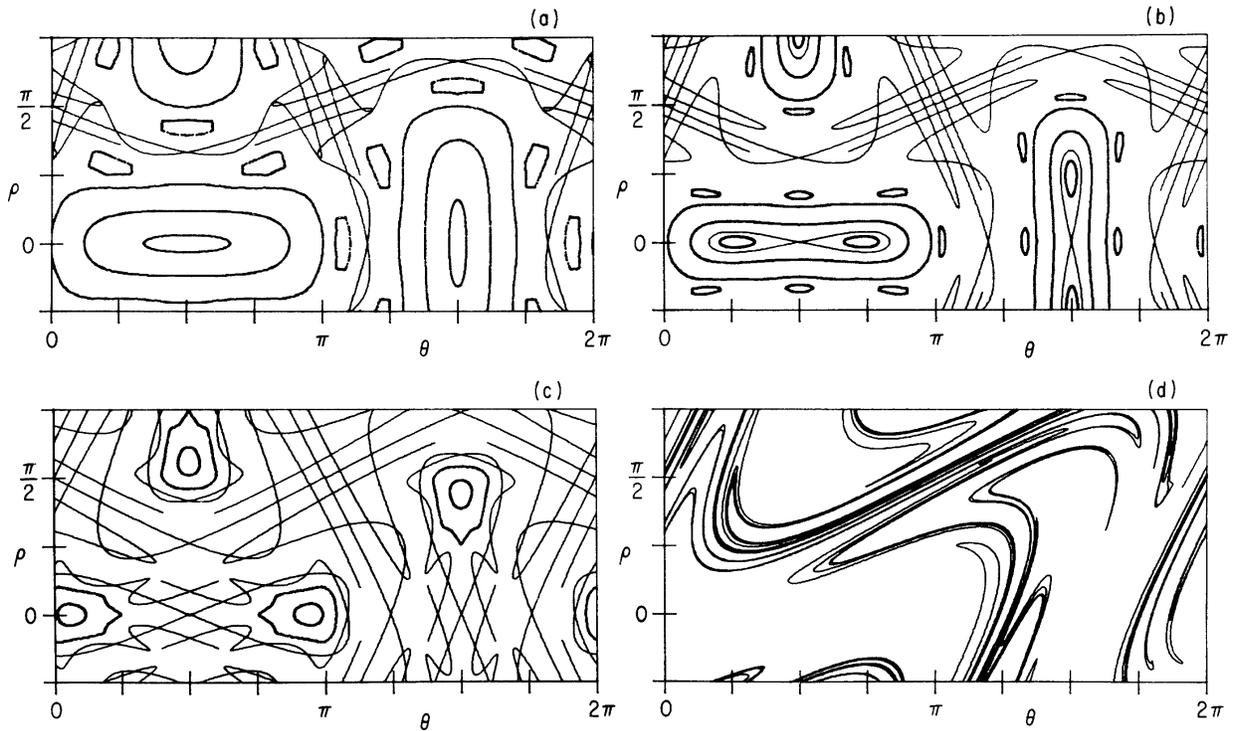
Eigencurves of the mapping, $T$, for $\d = 0$ and
(a) $A = 0.3$, (b) $A = 0.35$, (c) $A = 0.45$.  (d) The extension
of one of the eigencurves in (c) emanating from the hyperbolic
fixed point with reflection.
\Fig. 5.  (792075:0.3425)
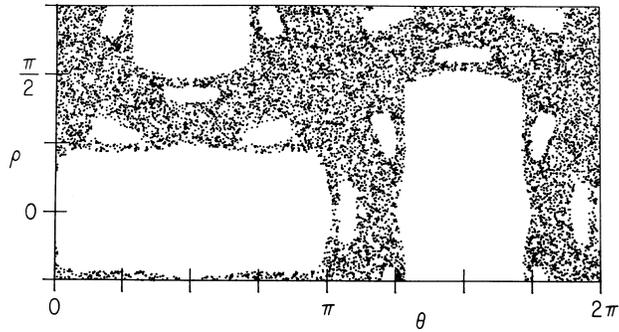
The iterates of a single point for $\d = 0$
and $A = 0.3$.  The point is chosen to start on one of the
eigencurves emanating from the hyperbolic fixed point and
$10\,000$ iterates are shown.  Compare with Fig. 4(a).
\Fig. 6.  (792073:0.5)
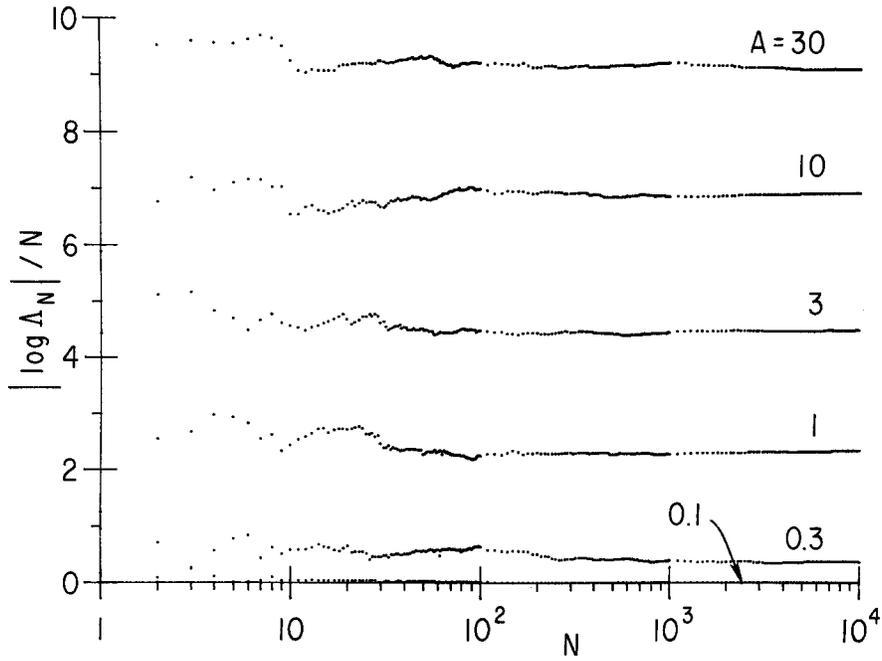
Plot of $\log\abs{\Lambda_N}/N$ as a function of $N$
for various values of $A$.  Here $\d = 0.23$ and the
reference trajectory was begun at $u_0 = 2$,
$v_0 = 6$.
\Fig. 7.  (792076:0.4)
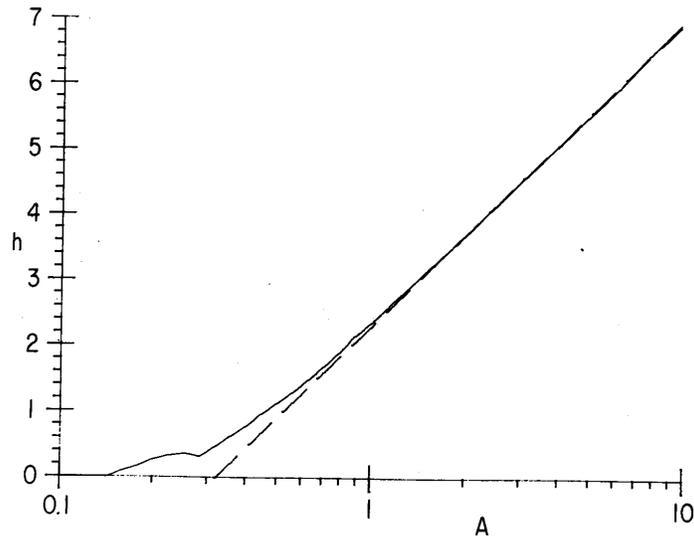
Plot of $h$ against $A$ for $\d = 0.23$,
$u_0 = 2$, $v_0 = 6$.  The dashed line is the
asymptotic result, (\hres).
\Fig. 8.  (792081:0.8)
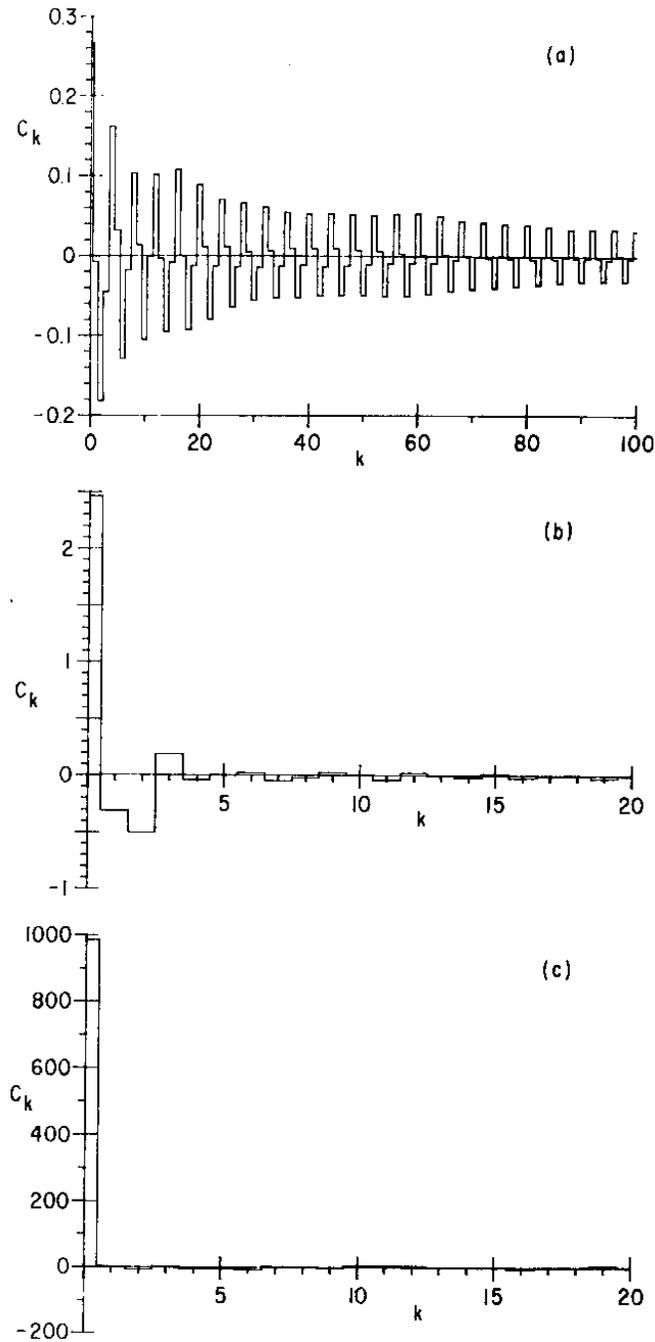
The correlation function, $C_k$, for $\d = 0.23$
and (a) $A = 0.2$, (b) $A = 0.5$, (c) $A = 10$.  $C_k$ was
computed using (\Ccomp) with $M = 10$ and $L = 5000$.  The
orbits were all chosen in the stochastic region.
\Fig. 9.  (792072:0.66667)
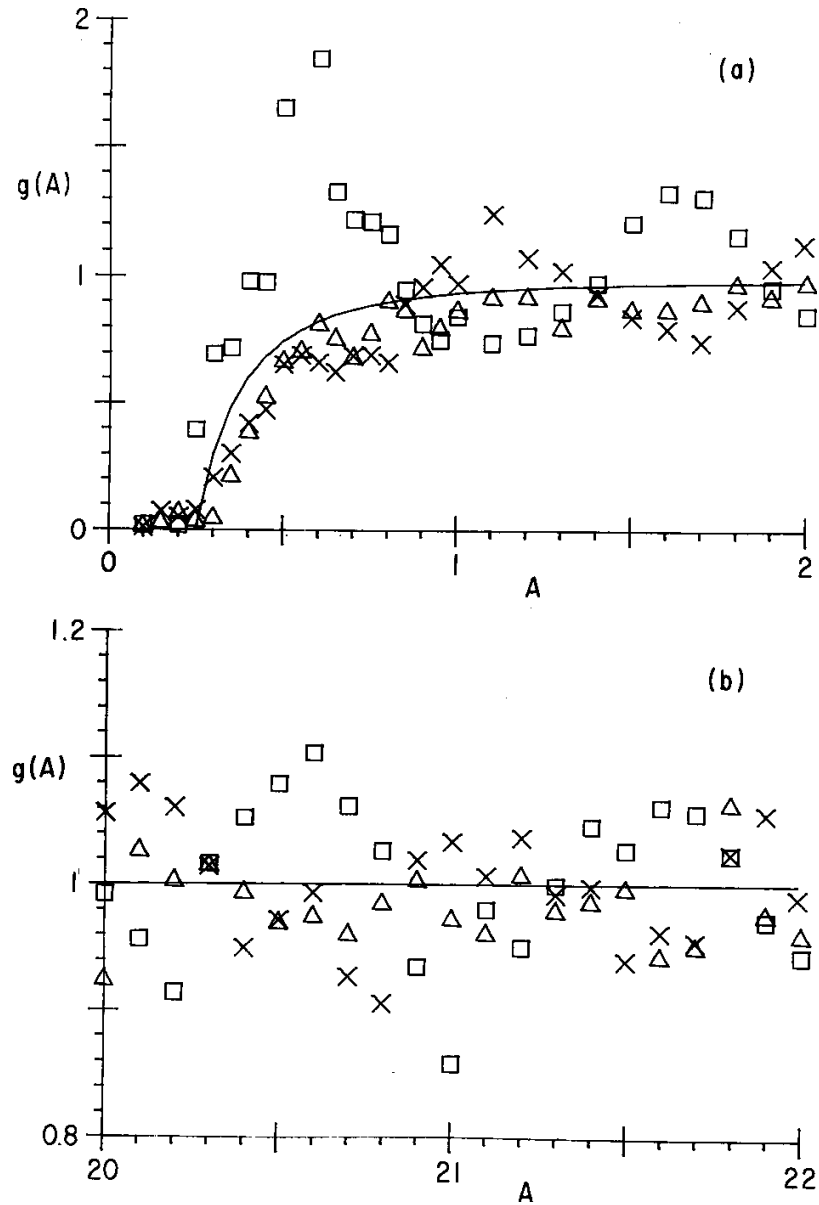
The function,
$g(A)$, against $A$ for $\d = 0.11$ (crosses),
0.23 (triangles), and 0.47 (squares).  (a) and (b) show
two different ranges of $A$.  The point
for $\d = 0.47$ and $A = 0.55$ is at $g(A) = 2.8$,
and so lies off the scale in (a).  The function, $g$, was evaluated
using (\Gdef) and (\Dcomp) with $M = 20$,
$L = 2500$, and (a) $K^\prime = 100$, $K = 150$,
(b) $K^\prime = 50$, $K = 100$.  The solid line gives the approximate
form for $g$, (\geq).
\Fig. 10.  (792077:0.5)
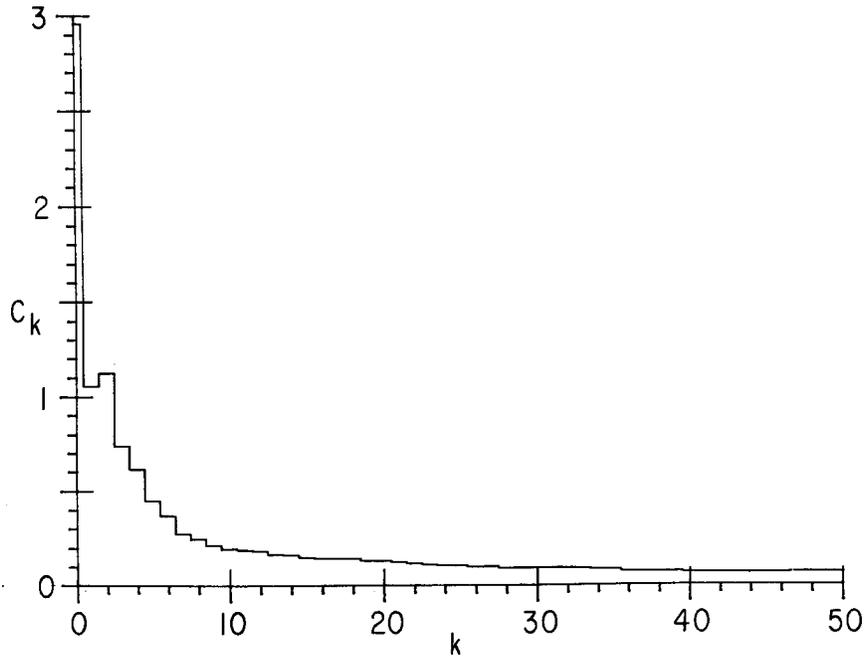
The correlation function $C_k$ for $\d = 0.47$,
$A = 0.55$, when an accelerator mode is present.  Here
we took $M = 100$ and $L = 5000$ in (\Ccomp).  The orbits
used to compute $C_k$ all lay outside the accelerator mode.
\Fig. 11.  (792089:0.66667)
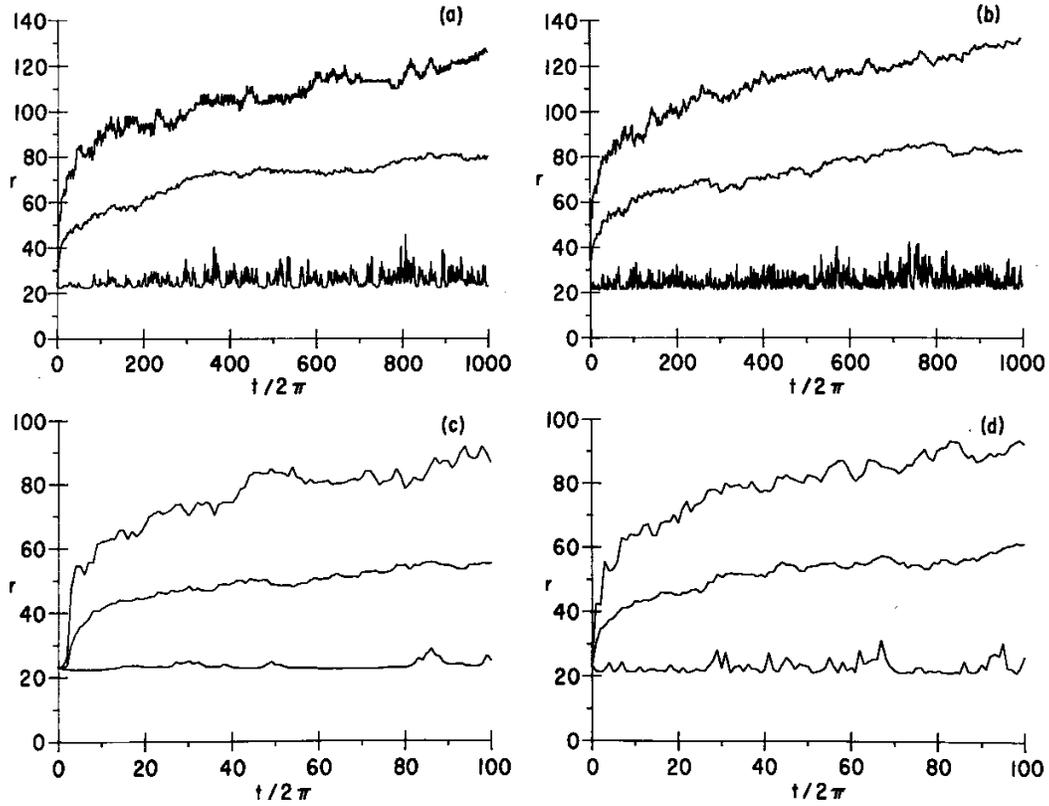
Maximum, minimum, and rms velocities of ions
as given by, (a) and (c), solution of the exact equations of motion,
(\lfl) (taken from I), and, (b) and (d), a Monte Carlo solution
of (\diffnorm).  In both cases the orbits of $N = 50$ particles
were followed with $\a = 20$, $\nu = 30.23$, and
initial velocity, $r_0 = 23$.  In (c) and (d) the time
scale is altered to show the short time behavior more
clearly.
\Fig. 12.  (792085:0.75)
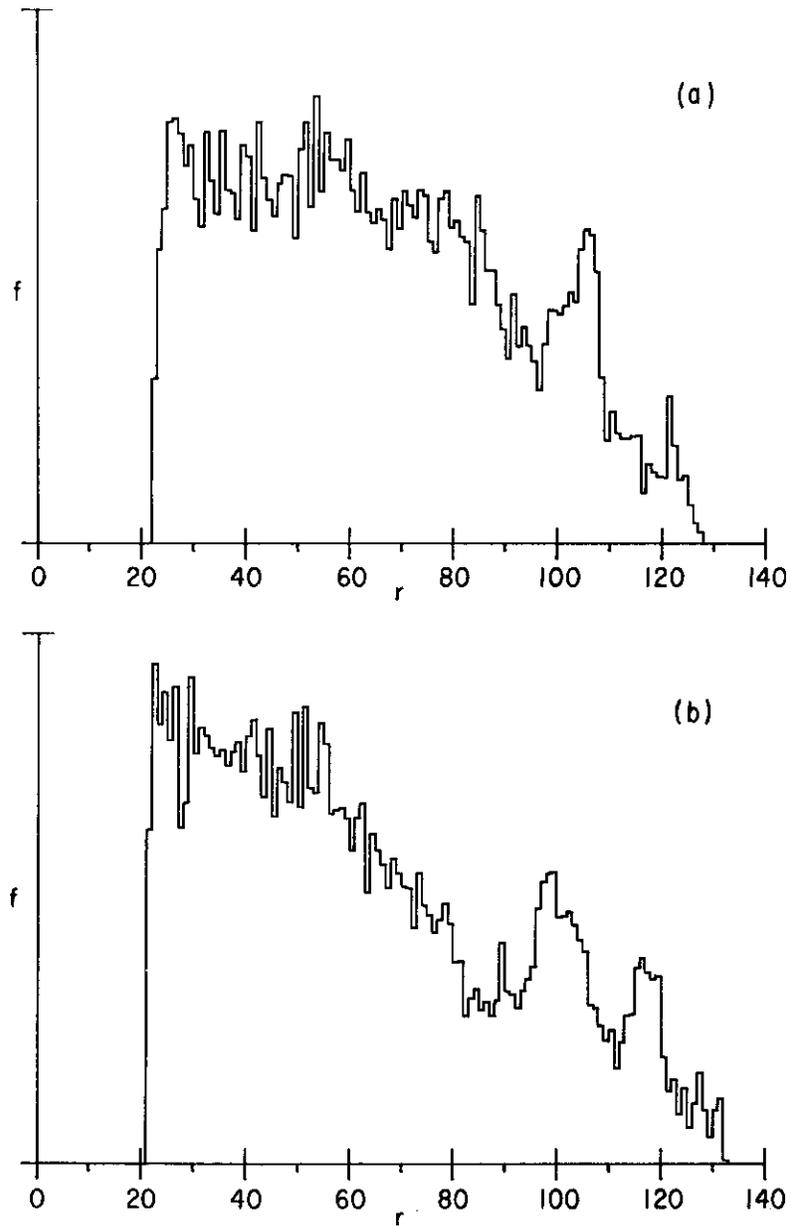
The perpendicular velocity distribution function
for the particles in Fig.\ 11 averaged over orbits 800--$1100$.  (a)
The distribution function obtained from the exact equation, (\lfl)
(taken from I).  (b) The distribution function obtained from
(\diffnorm).  In each case the normalization is such that
$\int2\pi r f\,dr =1$.
\Fig. 13.  (792079:0.5)
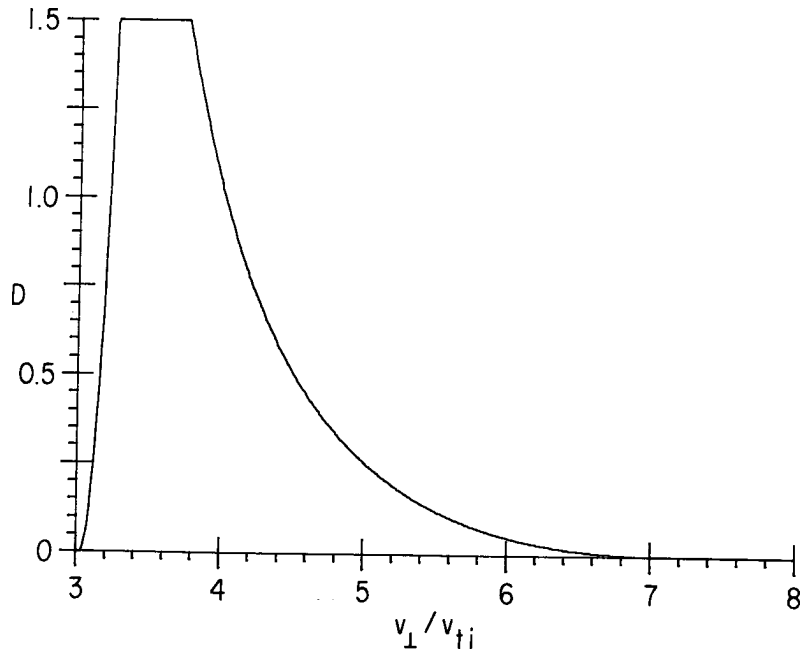
The diffusion coefficient $D$, (\Dun), as a function
of $v_\perp$ for the case discussed in Sec.\ XI\null.
\Fig. 14.  (792083:0.66667)
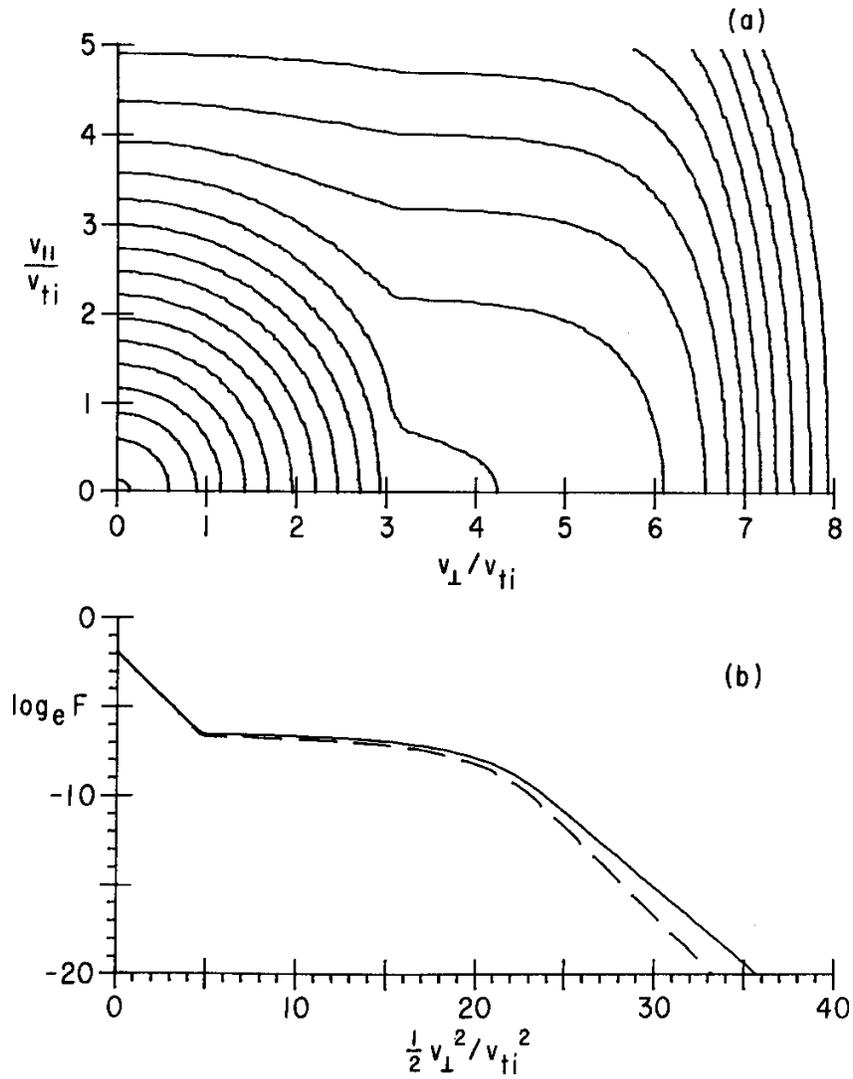
The steady-state solution for the distribution function.  (a)
The {\Twod} distribution function, $f$, as given by the
{\Twod} {\FP} equation.  (b) The {\Oned}
distribution function, $F(v_\perp)$, as given by integrating
the {\Twod} result over $v_\para$ (solid line) and solving
the {\Oned} {\FP} equation (dashed line).
\Fig. 15.  (792084:0.66667)
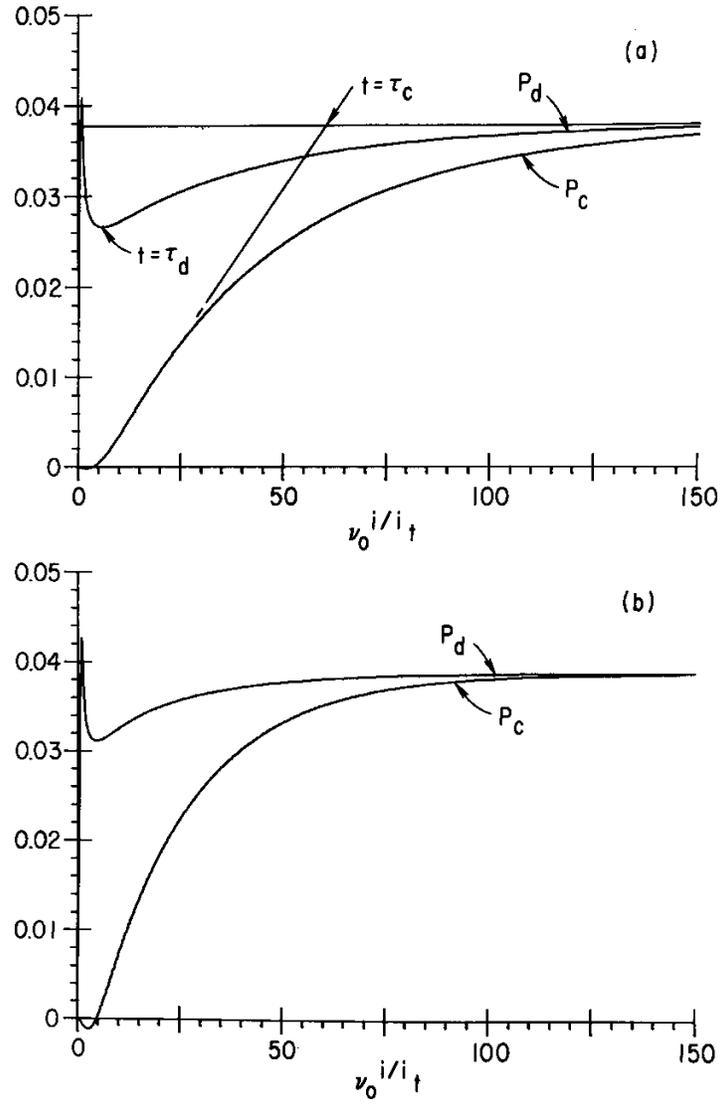
The power dissipated by the wave, $P_d$, and the
power lost to the background distributions, $P_c$, as a function
of time for the example given in Sec.\ XI\null.  The results
for the {\Twod} and {\Oned} cases are given in (a) and (b)
respectively.  The units of the vertical axes are
$m_i\vti^2n_0\nzero$.  The times $\tau_d$ and $\tau_c$ are
shown in (a).
\Fig. 16.  (792082:0.8)
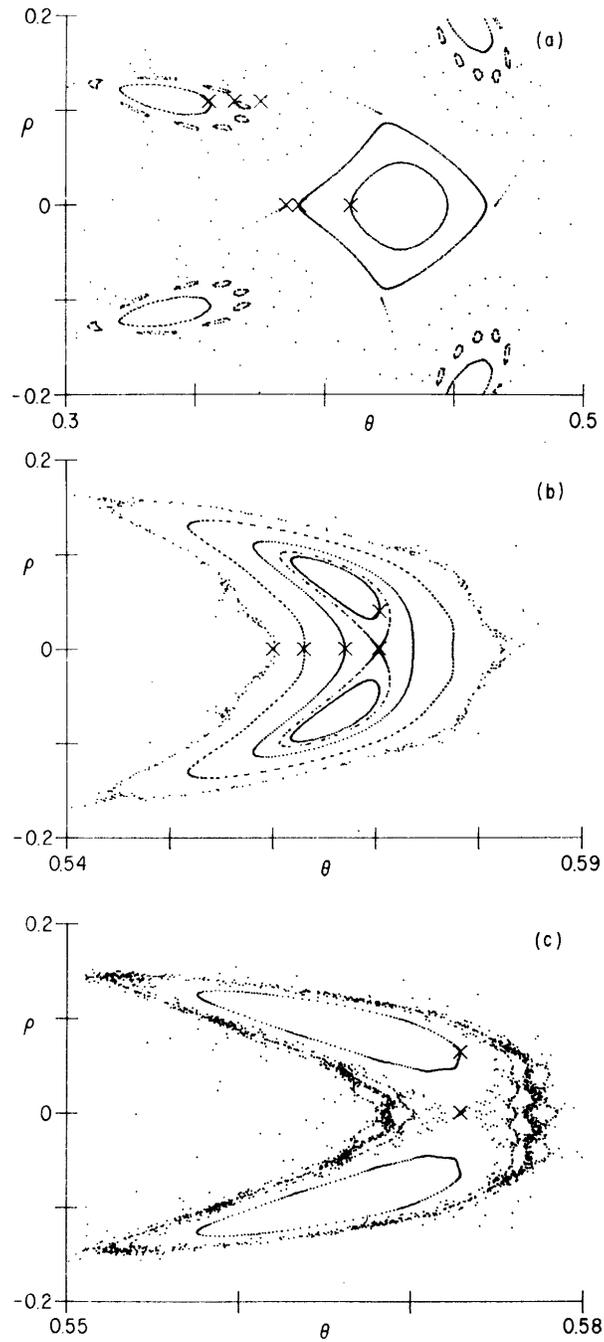
Development of the first-order accelerator
mode with $m = 1$, $n = 0$, and $\d = 0.5$.  The cases
shown are (a) $A = 0.55$, (b) $A = 0.594$,
and (c) $A = 0.595$.  For this mode $A_\l = 0.5$ and $A_u =
0.5927$.  The crosses show the starting positions of the particles.
\Fig. 17.  (792088:0.8)
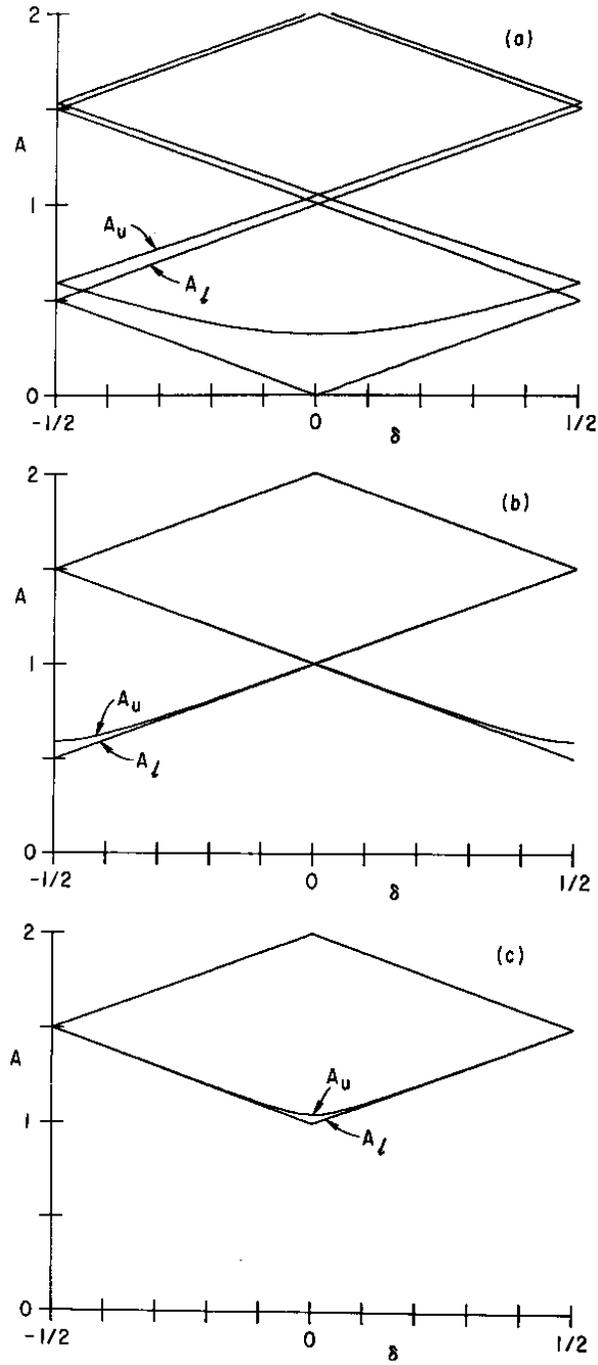
The range, $(A_\l, A_u)$, in which first-order
accelerator modes exist with (a) $s = 0$, (b)
$s = 1$, and (c) $s = 2$.
\Fig. 18.  (792074:0.5)
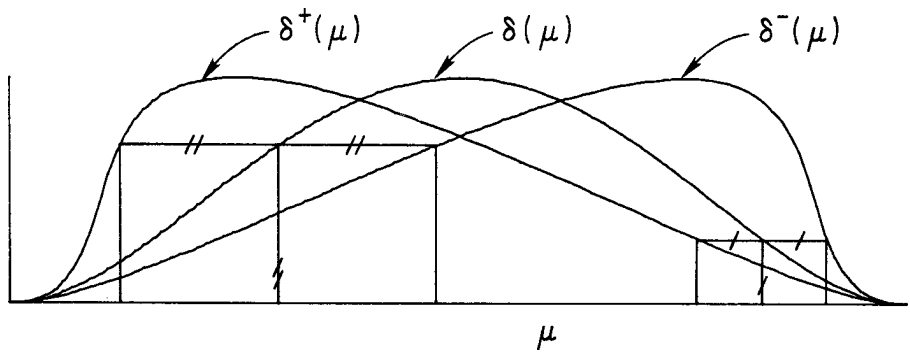
The graphical construction for $\dP(\mu)$
and $\dM(\mu)$ from $\d(\mu)$;  see (\stepeqp).
\bye